\documentclass[prb,aps,twocolumn,nopacs,superscriptaddress,nofootinbib,longbibliography]{revtex4-2}
\usepackage{amsmath}  \usepackage{amssymb}  \usepackage{amsfonts}  \usepackage{bm}  \usepackage{bbm}   \usepackage{braket}  \usepackage{color}  \usepackage{comment}  \usepackage{dcolumn}  \usepackage{enumerate}  
\usepackage{epsfig}  
\usepackage{gensymb}  \usepackage{graphicx}  \usepackage{indentfirst}  \usepackage{lmodern}  \usepackage{mathrsfs}  \usepackage{mathtools}  \usepackage{psfrag}  \usepackage{pst-all}  \usepackage{soul}  
\usepackage{xcolor}
\usepackage{float} %---APS---SI---arxiv
\usepackage[colorlinks,linkcolor=blue,citecolor=blue,urlcolor=blue,hyperindex,driverfallback=dvipdfm]{hyperref}  \usepackage[T1]{fontenc} %---APS---ACS---SI---arxiv

\def\ii{{\rm i}}  \def\ee{{\rm e}}
\def\me{m_{\rm e}}  \def\kB{{k_{\rm B}}}
\def\Ree{{\rm Re}}  \def\Imm{{\rm Im}}
    \def\Eb{{\bf E}}            \def\Hb{{\bf H}}      \def\jb{{\bf j}}              \def\Qb{{\bf Q}}    \def\Rb{{\bf R}}  \def\rb{{\bf r}}      \def\vb{{\bf v}} %--- bold vectors
    \def\zz{\hat{\bf z}}          
 \def\Ab{\hat{{\bf A}}}
\def\Hint{\hat{\mathcal{H}}^{\rm int}} 
\def\xit{\tilde{\xi}}

\begin{document}

\title{Electron decoherence in cylindrical holes and circular apertures}
\author{Cruz~I.~Velasco}%---
\email{cruzignaciovelasco@gmail.com}
\affiliation{ICFO-Institut de Ciencies Fotoniques, The Barcelona Institute of Science and Technology, 08860 Castelldefels (Barcelona), Spain}
\author{F.~Javier~Garc\'{\i}a~de~Abajo}%---
\email{javier.garciadeabajo@nanophotonics.es}
\affiliation{ICFO-Institut de Ciencies Fotoniques, The Barcelona Institute of Science and Technology, 08860 Castelldefels (Barcelona), Spain}
\affiliation{ICREA-Instituci\'o Catalana de Recerca i Estudis Avan\c{c}ats, Passeig Llu\'{\i}s Companys 23, 08010 Barcelona, Spain}

\begin{abstract}
The coherence of free electrons sets fundamental limits on the resolution and contrast of phase-sensitive electron microscopy because coherence is lost whenever the electron leaves a distinguishing excitation in its environment. We develop a quantitative theory of fast-electron decoherence in two canonical geometries: a cylindrical hole drilled through a realistic metal and a circular aperture in a thin perfectly conducting film. By combining an electromagnetic Green-tensor formulation with the fluctuation--dissipation theorem while fully retaining retardation, we obtain the decoherence probability and elastic phase as functions of the electron trajectory, temperature, and material response. For the cylindrical hole, we obtain a closed-form, azimuthally resolved expression that separates the dependences on path position, temperature, and conductivity. In the high-conductivity and high-temperature limits, this result reduces to a universal expression that is linear in temperature and independent of both conductivity and electron velocity. For the circular aperture, which we solve using a cylindrical-wave modal expansion, the energy-loss probability diverges as $1/\omega$ at low frequency, whereas the decoherence probability remains finite and reaches a maximum when the aperture radius is comparable to the relevant electromagnetic wavelength. Because the interaction is spatially localized, aperture-induced decoherence is generally weaker than that produced by a cylindrical hole, except for sufficiently large apertures at high temperature. Finally, we show that, for sufficiently large holes, decoherence can dominate the spatial broadening of a focused electron probe and must therefore be incorporated into the design of coherent electron-beam instruments.
\end{abstract}
\date{\today}
\maketitle
%\tableofcontents

% =========================================================
\section{Introduction}
\label{Introduction}

Electron microscopes use beams of fast electrons (e-beams) to characterize the structural and dynamical properties of materials down to the atomic scale. Their high spatial resolution stems from the short electron de Broglie wavelength, which decreases from tens to a few picometers as the velocity increases from $0.1\,c$ to $0.8\,c$, where $c$ is the speed of light in vacuum. Beyond structural imaging, the evanescent electromagnetic field carried by each electron spans a broad spectral range, enabling the excitation and detection of localized excitations, including dark modes, and providing access to spectroscopic information unavailable to optical methods~\cite{paper149}. A further class of techniques exploits the phase differences between coherent spatial components of the electron wave as an independent source of contrast. These include phase-contrast imaging~\cite{SAC19}, electron holography~\cite{LL02,WBT18}, electron interferometry~\cite{MD1956}, and electron diffraction~\cite{DG1927}.

This coherence is, however, readily degraded when the e-beam exchanges energy with its environment, including the specimen under investigation, nearby material structures, and the surrounding electromagnetic modes. Because different spatial components of the beam excite these environmental degrees of freedom with different amplitudes and phases, they become entangled with distinct, and generally distinguishable, environmental states. The initially coherent beam is thus converted into an incoherent mixture. Its ability to interfere is consequently reduced, and the overlap between spatially separated components contributes only a featureless background to the measured signal. We refer to this process as decoherence~\cite{APZ97,BHD96,H11_2,SL18,paper464}. In electron microscopy, decoherence limits the attainable contrast and spatial resolution. Excitations that carry away path information span a broad spectral range, from far-field photons to material excitations such as phonons, interband and intraband electronic transitions, and plasmons, as well as hybrid light--matter modes such as surface-plasmon polaritons~\cite{paper464}.

Previous studies of this mechanism have focused primarily on planar geometries. The canonical problem of a two-path electron moving parallel to a planar interface has been investigated extensively, both theoretically~\cite{F93,APZ97,BP01,MPV03,L04,HL06,M06_3,H11_2,SB12,H14_2,H19,K26} and experimentally~\cite{SH07,H10_2,BZB18, CB20,KRS20}. Among the experimental studies, Kerker \emph{et al.}~\cite{KRS20} reported agreement with the macroscopic quantum-electrodynamical treatment of Ref.~\cite{SB12}, while ruling out the models of Refs.~\cite{APZ97,M06_3,H11_2}. In a related study, decoherence induced by a conducting grating has been described in terms of the associated Smith--Purcell emission~\cite{AM08}. Relativistic formulations have subsequently been developed for trajectories near planar boundaries and edges~\cite{paper425}, including paths that penetrate the material or traverse a thin film~\cite{paper464}. Separately, fluctuating thermal (Johnson--Nyquist) magnetic fields generated by nearby conductors have been identified as a practical source of decoherence in electron optics~\cite{UMH13,UMZ15,MHS22}. The corresponding noise has been computed, within the quasistatic approximation, for conducting objects of simple geometry~\cite{LR08}. None of these treatments, however, considers a material boundary that curves around the electron trajectory while retaining the full retarded electromagnetic response. Existing calculations for cylindrical geometries are restricted to quasistatic magnetic noise~\cite{LR08,UMZ15}, even though retardation can be essential for an accurate description of the effect~\cite{paper040,H14_2,paper464}.

% Figure 1 ------------------------------------------------
\begin{figure}\centering\includegraphics[width=0.5\textwidth]{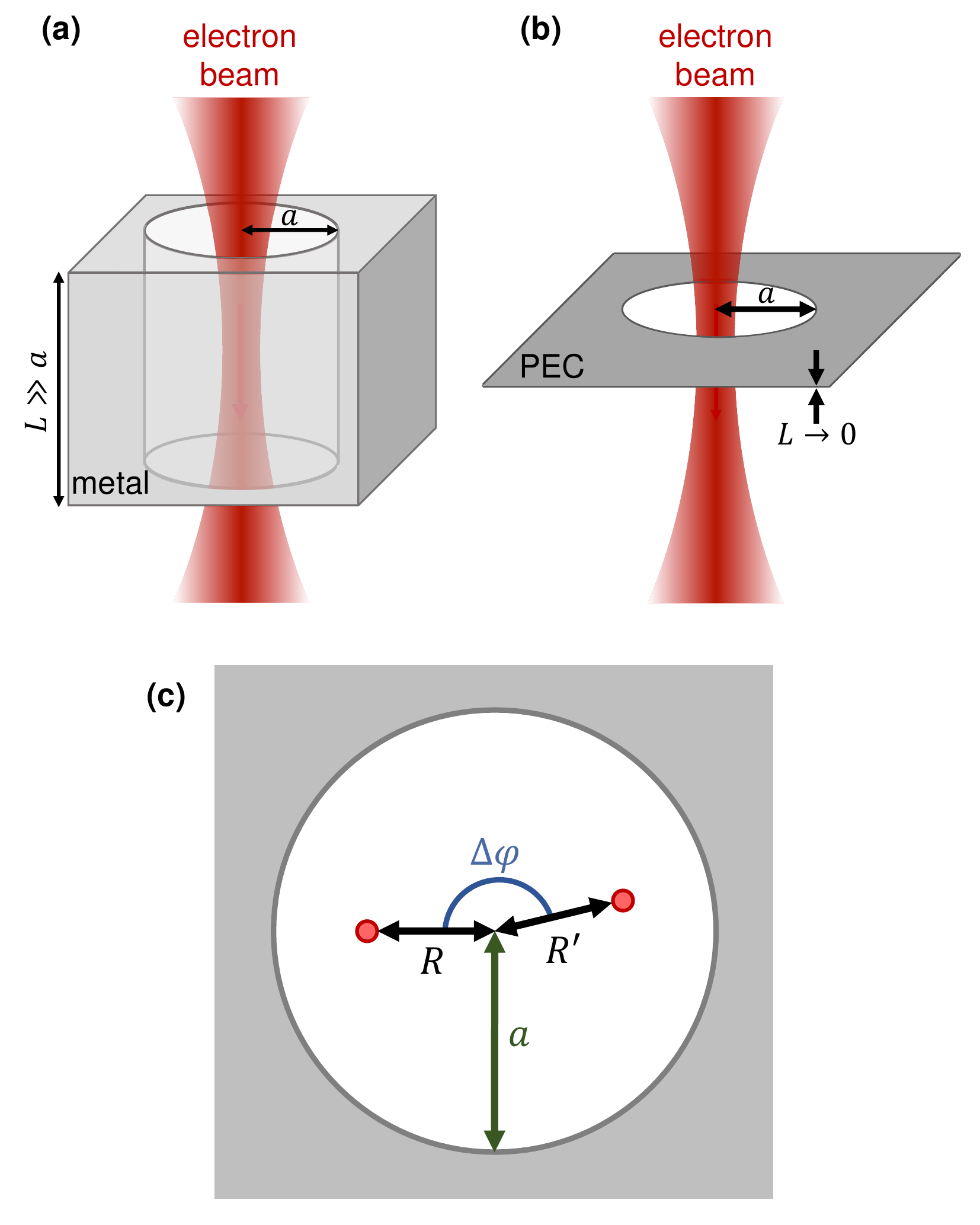}
\caption{\textbf{Physical systems under consideration}. \textbf{(a)}~Electron beam propagating through a cylindrical hole of radius $a$ and length $L\gg a$, drilled in a metallic material of permittivity $\epsilon(\omega)$. \textbf{(b)}~Electron beam traversing a circular aperture of radius $a$ in a thin perfectly conducting (PEC) film, corresponding to the limit $L\to0$. \textbf{(c)}~Top view of an e-beam prepared in a two-path superposition within a hole or aperture of radius $a$. The two point-like paths are located at transverse positions $\Rb=(R,\varphi)$ and $\Rb'=(R',\varphi')$, with relative azimuthal separation $\Delta\varphi=\varphi-\varphi'$.}
\label{Fig1}
\end{figure}

Building on a recent quantum formulation of free-electron decoherence in terms of the electromagnetic Green tensor and the fluctuation--dissipation theorem~\cite{paper425,paper464}, we develop a self-contained quantitative theory for electrons interacting with the two canonical geometries sketched in Fig.~\ref{Fig1}: a long cylindrical hole of radius $a$ drilled through a realistic metal, with interaction length $L\gg a$ (Sec.~\ref{Interaction}); and a circular aperture of the same radius in a thin perfectly conducting film, corresponding to the opposite limit $L\to0$ (Sec.~\ref{film}). These geometries bracket experimentally relevant situations and isolate two complementary mechanisms of coherence loss: coupling to confined modes along an extended channel and coupling to diffraction-radiation modes during passage through a single aperture.

We first summarize the general formalism governing the evolution of the reduced density matrix in Sec.~\ref{Generaltheory}. For the cylindrical hole, discussed in Sec.~\ref{Interaction}, we obtain an analytical expression for the decoherence probability that separates the dependence on trajectory position, temperature, and conductivity, complementing the relativistic energy-loss analysis of cylindrical waveguides and holes presented in Ref.~\cite{paper448}. In the experimentally relevant high-conductivity and high-temperature regime, this result reduces to a universal expression that is linear in temperature and independent of both conductivity and electron velocity, consistent with the weak material dependence observed experimentally~\cite{UMZ15}. Then, we compare the results for the cylindrical hole with the archetypal planar slab where both a theoretical model and experimental data are available. For the circular aperture considered in Sec.~\ref{film}, translational symmetry is absent, and the response cannot be described by a single reflection coefficient. We therefore develop a rigorous modal expansion and solve it numerically. The corresponding energy-loss probability diverges as $1/\omega$ at low frequency, as anticipated from the perfectly conducting half-plane geometry~\cite{paper425}, whereas the decoherence probability remains finite and reaches a maximum near $\omega a/c\sim1$. Finally, in Sec.~\ref{imaging}, we quantify the impact of these effects on interference and image formation, showing that decoherence becomes the dominant source of focused-probe degradation for sufficiently large holes. Our conclusions are summarized in Sec.~\ref{conclusion}.

% =========================================================
\section{Electron decoherence mediated by material electromagnetic response}
\label{Generaltheory}

We begin by summarizing the formalism presented in Refs.~\cite{paper425,paper464}, which we apply below to the two configurations shown in Fig.~\ref{Fig1}. We consider an electron moving with velocity $\vb$ in the presence of a material structure. Its state is described by the reduced density matrix $\rho(\rb,\rb',t)$ [e.g., $\psi(\rb,t)\psi^*(\rb',t)$ for a pure state], obtained by tracing the joint electron--environment density matrix over the environmental degrees of freedom. We assume that the electron and its environment are initially uncorrelated. The environment comprises the material structure and the electromagnetic modes mediating the electron--material interaction, and is taken to be in thermal equilibrium at temperature $T$. Each mode of frequency $\omega$ is initially populated according to the Bose--Einstein distribution $n_T(\omega) = \big(\ee^{\hbar\omega/\kB T}-1\big)^{-1}$.

The interaction is introduced through the minimal-coupling Hamiltonian $\Hint(\rb) = e(\vb/c) \cdot \Ab(\rb)$, where $\Ab(\rb)$ is the electromagnetic vector-potential operator. We neglect terms proportional to $\hat{A}^2$, assuming that no external illumination is present. We also adopt the nonrecoil approximation, taking both the energy spread of the incident electron and the energy exchanged with the material to be small compared with the mean kinetic energy $\mathcal{E}_0=\me c^2(\gamma-1)$, with $\gamma=1/\sqrt{1-(v/c)^2}$. The electron velocity $\vb$ may therefore be treated as constant throughout the interaction.

The interaction entangles the electron with its environment, partially suppressing its coherence and transforming the incident reduced density matrix $\rho_i(\rb,\rb')$ into
\begin{align}\label{evol}%--
\rho(\rb,\rb') = \ee^{-P(\rb,\rb')+\ii \chi(\rb,\rb')} \rho_i(\rb,\rb')
\end{align}
after the interaction. Here, $P(\rb,\rb')$ is a non-negative decoherence exponent that is referred to as decoherence probability throughout the paper. It satisfies $P(\rb,\rb)=0$, so that the electron probability density $\rho(\rb,\rb)$ is locally conserved. The phase $\chi(\rb,\rb')$, which arises from the elastic, energy-conserving component of the interaction, likewise vanishes for $\rb=\rb'$.

Both $P$ and $\chi$ can be expressed in terms of the electromagnetic Green tensor $G(\rb,\rb',\omega)$. Using the fluctuation--dissipation theorem, the electromagnetic-field correlations are related directly to $\Imm\{G(\rb,\rb',\omega)\}$. The Green tensor satisfies
\begin{align}\label{greentens}%-- 
\nabla \times \nabla \times G(\rb,\rb',\omega) - \frac{\omega^2}{c^2}\,\epsilon(\rb,\omega)\cdot G(\rb,\rb',\omega) 
\\\nonumber%--
= -\frac{1}{c^2} \delta(\rb-\rb'),
\end{align}
where $\epsilon(\rb,\omega)$ is a frequency- and position-dependent local dielectric tensor. The extension to nonlocal media is straightforward \cite{paper357}. The material properties therefore enter through $\epsilon(\rb,\omega)$, while the geometry is encoded in the boundary conditions imposed on the Green tensor.

The decoherence probability and elastic phase read
\begin{widetext}%--
\begin{subequations}%--
\begin{align}\label{decoP}%--
P(\rb,\rb') &= \frac{1}{2} \int_0^{\infty} d\omega [2n_T(\omega)+1] \left[ \Gamma(\rb,\rb,\omega) + \Gamma(\rb',\rb',\omega) - 2 \Gamma(\rb,\rb',\omega) \right]
\\\label{phase}%--
\chi(\rb,\rb') &= \frac{2e^2}{\hbar}\int_0^{\infty} d\omega \int_{-\infty}^{\infty} dz''\int_{-\infty}^{\infty} dz''' \bigg\{ 2\sin\left[ \frac{\omega}{v} (z-z'-z''+z''')\right] \Imm\{G_{zz}(\Rb,z'',\Rb',z''',\omega)\} \\ \nonumber
&\quad\quad+\cos\left[\frac{\omega}{v}(z''-z''')\right] \Ree\{G_{zz}(\Rb',z'',\Rb',z''',\omega)-G_{zz}(\Rb,z'',\Rb,z''',\omega)\}\bigg\},
\end{align}    
\end{subequations}
where $\Rb = (x,y)$ is the component of $\rb$ transverse to the e-beam direction, and 
\begin{align}\label{eels}%--
\Gamma(\rb,\rb',\omega)
= \frac{4e^2}{\hbar} \int_{-\infty}^{\infty} dz'' \int_{-\infty}^{\infty}dz''' \cos\bigg[\frac{\omega}{v}(z-z'-z''+z''') \bigg]\, \Imm\{-G_{zz}(\Rb,z'',\Rb',z''',\omega)\}
\end{align}
\end{widetext}%--
is a nonlocal energy-loss kernel correlating two lateral components of the electron wave. Because we are interested in the influence of $P$ and $\chi$ on interferometric measurements and image formation (see Sec.~\ref{imaging}), we set $z=z'$ and retain the reduced density matrix only as a function of the transverse coordinates $\Rb$ and $\Rb'$. For notational simplicity, we therefore suppress the $z$-dependence and write $P(\Rb,\Rb')$ and $\chi(\Rb,\Rb')$.

The loss probability can be written in an equivalent and physically more transparent form in terms of the field induced by the electron along its own trajectory~\cite{paper464}. Within the nonrecoil approximation, the electron is represented as a classical line current $\jb(\rb,\omega)=-e\,\zz\,\ee^{\ii\omega z/v}\,\delta(\Rb-\Rb')$, located at the transverse position $\Rb'$. The corresponding induced electric field is obtained from the Green tensor as
\begin{align}\label{Efield}%--
\Eb(\rb,\Rb',\omega)=4\pi\ii e\omega\int_{-\infty}^{\infty}dz'\,\ee^{\ii\omega z'/v}\,G(\rb,\rb',\omega)\cdot\zz.
\end{align}
The nonlocal energy-loss probability can then be expressed as
\begin{align}\label{Gammasym}%--
\Gamma(\Rb,\Rb',\omega)=\frac{1}{2}\big[\tilde\Gamma(\Rb,\Rb',\omega)+\tilde\Gamma(\Rb',\Rb,\omega)\big],
\end{align}
with
\begin{align}\label{Gammatilde}%--
\tilde\Gamma(\Rb,\Rb',\omega)=\frac{e}{\pi\hbar\omega}\int_{-\infty}^{\infty}dz\,\Ree\big\{\ee^{-\ii\omega z/v}E_z(\rb,\Rb',\omega)\big\}.
\end{align}
This form makes explicit that $\Gamma(\Rb,\Rb',\omega)$ describes the exchange of an energy quantum $\hbar\omega$ between two lateral components of the electron wave located at transverse positions $\Rb$ and $\Rb'$. In Sec.~\ref{film}, we use Eq.~(\ref{Gammatilde}), because $E_z$ is obtained numerically rather than from a closed-form expression for the Green tensor.

% =========================================================
\section{Decoherence by an infinite cylindrical hole}
\label{Interaction}

We consider an electron moving with velocity $\vb$ inside a cylindrical hole of radius $a$, along a trajectory parallel to the hole axis. The hole has length $L\gg a$ and is drilled through a homogeneous and isotropic material characterized by the frequency-dependent permittivity $\epsilon(\omega)$. We neglect effects associated with the entrance and exit edges, whose contribution is analyzed separately in Sec.~\ref{film}.

% Figure 2 ------------------------------------------------
\begin{figure*}\centering\includegraphics[width=1.00\textwidth]{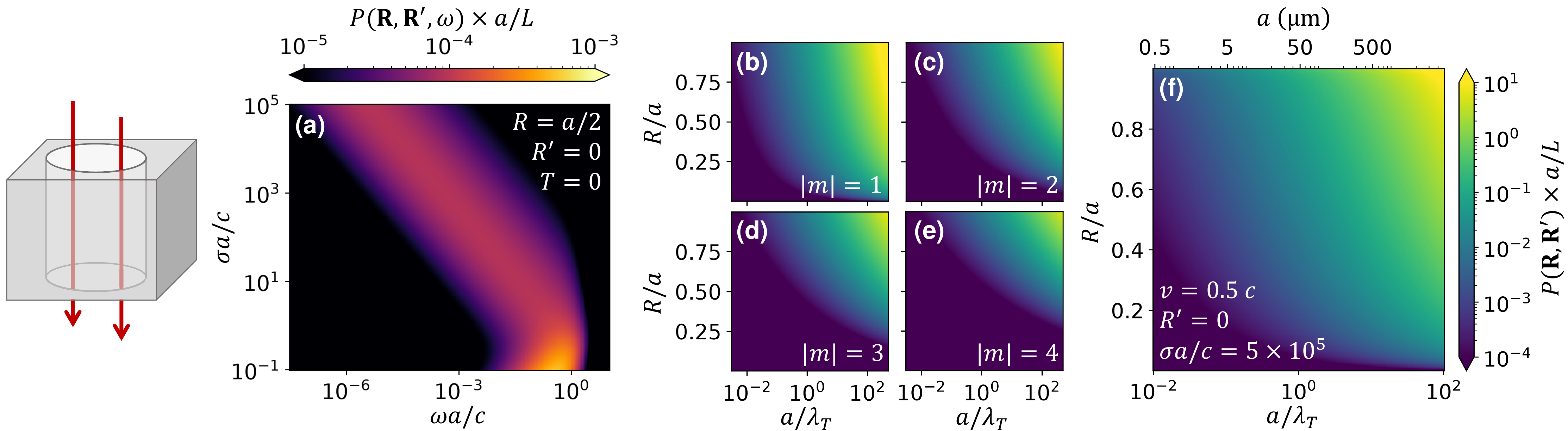}
\caption{\textbf{Decoherence in a cylindrical hole}. \textbf{(a)}~Spectral density of the decoherence probability for a two-path electron traversing a cylindrical hole drilled through a metallic slab of dc conductivity $\sigma$ [see Fig.~\ref{Fig1}(a,c)], plotted as a function of frequency $\omega$ and $\sigma$, both normalized by $c/a$, for $R=a/2$, $R'=0$, and $T=0$. \textbf{(b)}~Contribution of the azimuthal modes with $|m|=1$ to the decoherence probability for $v=0.5\,c$, $R'=0$, and $\sigma a/c=5\times10^{5}$, shown as a function of the hole radius normalized to the thermal wavelength $\lambda_T=2\pi\hbar c/k_BT$. \textbf{(c--e)}~Same as (b), but for $|m|=2$, $|m|=3$, and $|m|=4$, respectively. \textbf{(f)}~Total decoherence probability obtained by summing over all $m$. The upper horizontal axis gives the hole radius for a temperature $T=300$~K ($\lambda_T\approx50\,\mu$m). In all panels, the decoherence probability is normalized by the hole length-to-radius ratio $L/a$.}
\label{Fig2}
\end{figure*}

The Green tensor for this system is derived in Appendix~\ref{apdxeels} by imposing the boundary conditions of an infinite cylindrical structure. When inserted in Eq.~(\ref{eels}) it yields the corresponding generalized loss probability 
\begin{align}\label{decoGamma}%--
\Gamma(\Rb,\Rb',\omega) &= \frac{e^2 L}{\hbar v^2\gamma^2} \sum_{m=-\infty}^{\infty} \cos\big[ m (\varphi-\varphi') \big]
\\\nonumber%--
&\times I_m\bigg(\frac{\omega R}{v\gamma}\bigg) I_m\bigg(\frac{\omega R'}{v\gamma}\bigg)\; \Ree\big\{ (-1)^{m+1} r^m_{pp}\big\}.
\end{align}
Here, we use cylindrical coordinates $\rb=(R,\varphi,z)$, $I_m(x)$ is a modified Bessel function, and $r_{\rm pp}^m$ is the internal reflection coefficient for ${\rm p}$-polarized outgoing cylindrical waves inside the hole. The dielectric properties of the surrounding medium are encapsulated in $r_{\rm pp}^m$, which is obtained by solving the electromagnetic boundary conditions at the hole surface $R=a$. For a hole in a homogeneous isotropic material, we find (see Appendix~\ref{refcoeff})
\begin{align}\label{rppm}%--
r_{\rm pp}^m = \frac{2\ii}{\pi}(-1)^m \bigg\{ \frac{K_m(\xi)}{I_m(\xi)} + \frac{r_m}{I_m^2(\xi)}\bigg\},
\end{align}
where
\begin{align}\label{rmsigma}%--
&r_m \\ \nonumber
&= \frac{-\xit/\xi}{\xit\, f_m(\xi)-\epsilon\,\xi\, g_m(\xit)-\dfrac{m^2c^2}{v^2} \dfrac{\big[\xi/\xit-\xit/\xi\big]^2}{\xit f_m(\xi)-\xi g_m(\xit)}}, 
\end{align}
with
\begin{align}\nonumber%--
&\xi = a\omega/v\gamma,
\\\nonumber%--
&\xit = \xi \gamma \sqrt{1-(v/c)^2\epsilon},
\\\nonumber%--
&f_m(\xi) = I_m'(\xi)/I_m(\xi),
\\\nonumber%--
&g_m(\xit) = K_m'(\xit)/K_m(\xit), 
\end{align}
and $K_m(x)$ is a modified Bessel function of the second kind.

In electron microscopy, one is typically concerned with hole diameters above a few microns, for which the electron--hole interaction is dominated by low frequencies in the mid-infrared and below. For a conductive medium outside the hole, the dielectric function is then approximately described by the Drude model~\cite{DG02} as $\epsilon(\omega)=1+4\pi\ii\sigma/\omega$, where $\sigma$ is the dc conductivity. Combining these results, we obtain
\begin{align}\label{decoPres}%--
P(\Rb,\Rb')& = \frac{\alpha c L}{\pi v\gamma a} \int_0^\infty d\xi \coth\bigg(\frac{ v\gamma \lambda_T}{4\pi a c}\,\xi\bigg)
\\\nonumber%--
&\times\sum_{m=-\infty}^\infty \Imm\{r_m(v,\sigma,\xi)\} A_m(R,R',\Delta\varphi,\xi)
\end{align}    
for the decoherence probability, where $\alpha=e^2/\hbar c$ is the fine-structure constant, $\Delta\varphi=\varphi-\varphi'$ is the angular separation, and $\lambda_T = 2\pi\hbar c/k_B T$ is the thermal wavelength. The factor
\begin{align}\nonumber%--
A_m(R,R',&\Delta\varphi,\xi) = \frac{1}{I_m^2(\xi)}\Bigg[I_m^2\left(\frac{\xi R}{a}\right)+ I_m^2\left(\frac{\xi R'}{a}\right)
\\\label{Am}%--
&-2\cos(m\Delta\varphi)\, I_m\left(\frac{\xi R}{a}\right)\, I_m\left(\frac{\xi R'}{a}\right)\Bigg]
\end{align}
fully encodes the path-position dependence, while
\begin{align}\nonumber%--
\frac{2\, \Imm\{r_m(v,\sigma,\xi)\}}{\pi I^2_m(\xi)} = \Ree\big\{(-1)^{m+1} r_{\rm pp}^m\big\}
\end{align}
contains the dependence on the surrounding material. Here, $r_m$ is defined in Eq.~(\ref{rmsigma}) and we have dropped the first term in the curly brackets in Eq.~(\ref{rppm}) because it is purely imaginary. 

The elastic phase follows from an analogous procedure and reads
\begin{align}\label{chires}%--
\chi(\Rb,\Rb') &= \frac{\alpha c L}{\pi v \gamma a} \int_0^\infty d\xi \\ \nonumber
&\times \sum_{m = -\infty}^{\infty} \frac{u_m(v,\sigma,\xi)}{I_m^2(\xi)} \left[I^2_m\left(\frac{\xi R}{a}\right)-I^2_m\left(\frac{\xi R'}{a}\right)\right],
\end{align}
where again the properties of the medium enter through the reflection coefficient as
\begin{align}\nonumber%--
\Imm\big\{(-1)^{m+1}r_{\rm pp}^m\big\} = \frac{2u_m(v,\sigma,\xi)}{\pi I^2_m(\xi)}.
\end{align}
We note that Eqs.~(\ref{decoPres}) and~(\ref{chires}) apply to arbitrary transverse e-beam profiles. Below, we analyze the case of an electron prepared in a two-path superposition state, schematized in Fig.~\ref{Fig1}(c), with each path corresponding to a focused beam of width $w\ll a$, allowing us to interpret $P(\Rb,\Rb')$ as the interpath decoherence probability in the $w\ll a$ limit. In Sec.~\ref{imaging}, we turn to extended beams.

%----------------------------------------------------------
\subsection{Finite-conductivity results} \label{finitecond}

We begin by examining the spectral density of the decoherence probability, defined as $P(\Rb,\Rb',\omega)$ such that $P(\Rb,\Rb') =\int_0^{\infty} d\omega P(\Rb,\Rb',\omega)$. In Fig.~\ref{Fig2}(a) we show this quantity for $R=a/2$, $R'=0$, and $T=0$, as a function of dimensionless frequency $\omega a/c$ and conductivity $\sigma a/c$. The spectral density exhibits a maximum whose position depends strongly on the dc conductivity of the material: for $\sigma a/c\gg1$, the maximum occurs at $\omega a/c\sim(\sigma a/c)^{-1}$, coinciding with the regime where the field penetration depth, $\delta_\omega = c/\sqrt{2\pi\sigma\omega}$, is comparable to the size of the hole. Therefore, this maximum shifts toward lower frequencies as the conductivity increases. When $\sigma a /c\lesssim 1$, this condition would place the maximum beyond $\omega\sim c/a$. However, near this spectral region, the transverse extent of the electron's evanescent field, given by $v\gamma/\omega$, falls below its distance to the wall and the coupling is exponentially suppressed [see Eq.~(\ref{Am})], so the maximum remains near $\omega a/c\sim 1$ independently of $\sigma$. In this regime, $\omega\sim c/a$ also becomes comparable to $4\pi\sigma$, such that $\epsilon(\omega)$ is of order unity and the material stops behaving as a good conductor, becoming strongly absorbing instead. This behavior increases the spectral density compared with the scenario in which the material has high conductivity.

% Figure 3 ------------------------------------------------
\begin{figure}\centering\includegraphics[width = 0.5\textwidth]{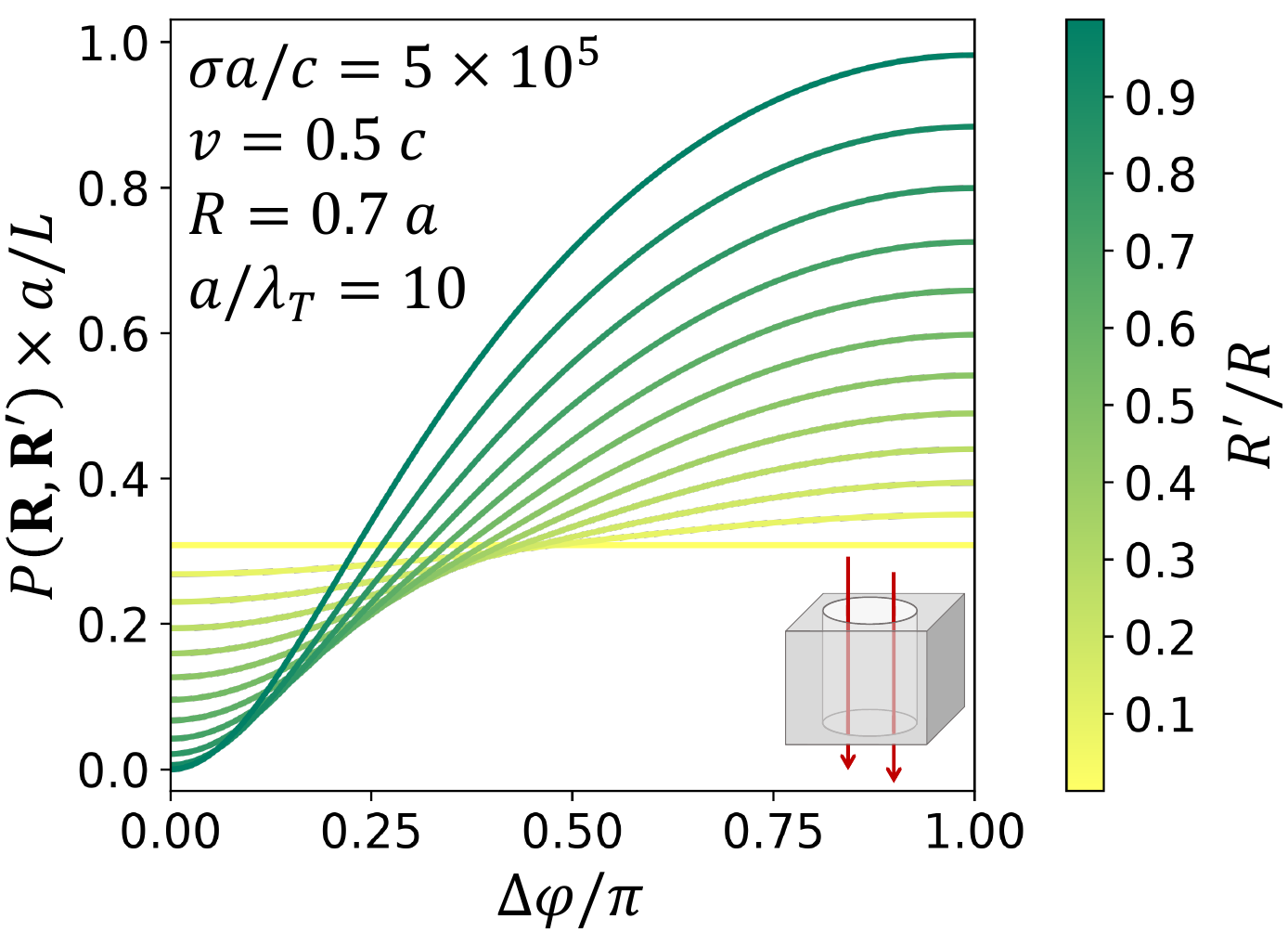}
\caption{\textbf{Angular dependence of the decoherence probability}. Decoherence probability, normalized by the hole length-to-radius ratio $L/a$ (i.e., $P\times a/L$), for electron paths located at polar coordinates $(R,\varphi)$ and $(R',\varphi')$ [see Fig.~\ref{Fig1}(c)], plotted as a function of their relative azimuthal separation $\Delta\varphi=\varphi-\varphi'$. Results are shown for $v=0.5\,c$, $\sigma a/c=5\times10^5$, $a/\lambda_T = 10$, $R=0.7\,a$, and several values of the $R'/R$ ratio (see color scale).}
\label{Fig3}
\end{figure}

The full probability, obtained after integrating over frequency, is shown in Fig.~\ref{Fig2}(b--f) as a function of the radial position $R$ of one path, with the other fixed at the axis ($R'=0$), and of temperature, the latter represented through the inverse thermal wavelength. We keep the velocity and angular positions as before and fix $\sigma a/c=5\times10^{5}$, which for typical metallic conductivities ($\sigma/c\sim1\,{\rm nm}^{-1}$~\cite{HLB17}) corresponds to a sub-millimeter hole radius, $a\sim0.5$~mm. The probability is normalized by the geometric ratio $L/a$. In this configuration, the decoherence is dominated by the low-order modes with $|m|\geq1$. In contrast, the contribution of the $m=0$ mode is negligible, because for a path on the axis, its position factor $A_0$ [Eq.~(\ref{Am})] vanishes as $\xi^4$ at low frequency, precisely the range that carries the spectral weight at high conductivity (see below). Most of $P$ originates from the $|m|=1$ term while the electron trajectories remain far from the walls, with higher-order terms contributing only as the paths approach the walls. Decoherence increases with temperature, because modes with $\hbar \omega \lesssim \kB T$ become thermally populated according to the Bose--Einstein distribution and thus their interaction with the beam becomes stimulated. It also increases with path separation: when the two paths are close, they become nearly indistinguishable to the modes inside the hole, and the cross term $\Gamma(\Rb,\Rb',\omega)$ nearly cancels the direct terms $\Gamma(\Rb,\Rb,\omega)$ and $\Gamma(\Rb',\Rb',\omega)$, yielding a small decoherence probability. This is corroborated by the angular dependence in Fig.~\ref{Fig3}, obtained at fixed $R$ by varying $\Delta\varphi$ for several $R'/R$ ratios. Here, $P$ is negligible for small angular separations and grows with $\Delta\varphi$, with the variation becoming less pronounced as the difference between $R$ and $R'$ increases. We remark that in Fig.~\ref{Fig3} we have fixed the ratio of the radius to the thermal wavelength $a/\lambda_T=10$. In Supplementary Fig.~\ref{FigS1}, we examine different values of this ratio. Although the oscillation amplitude varies significantly, the shapes of the curves remain largely unchanged.

% Figure 4 ------------------------------------------------
\begin{figure*}[th!]\centering\includegraphics[width = 0.9\textwidth]{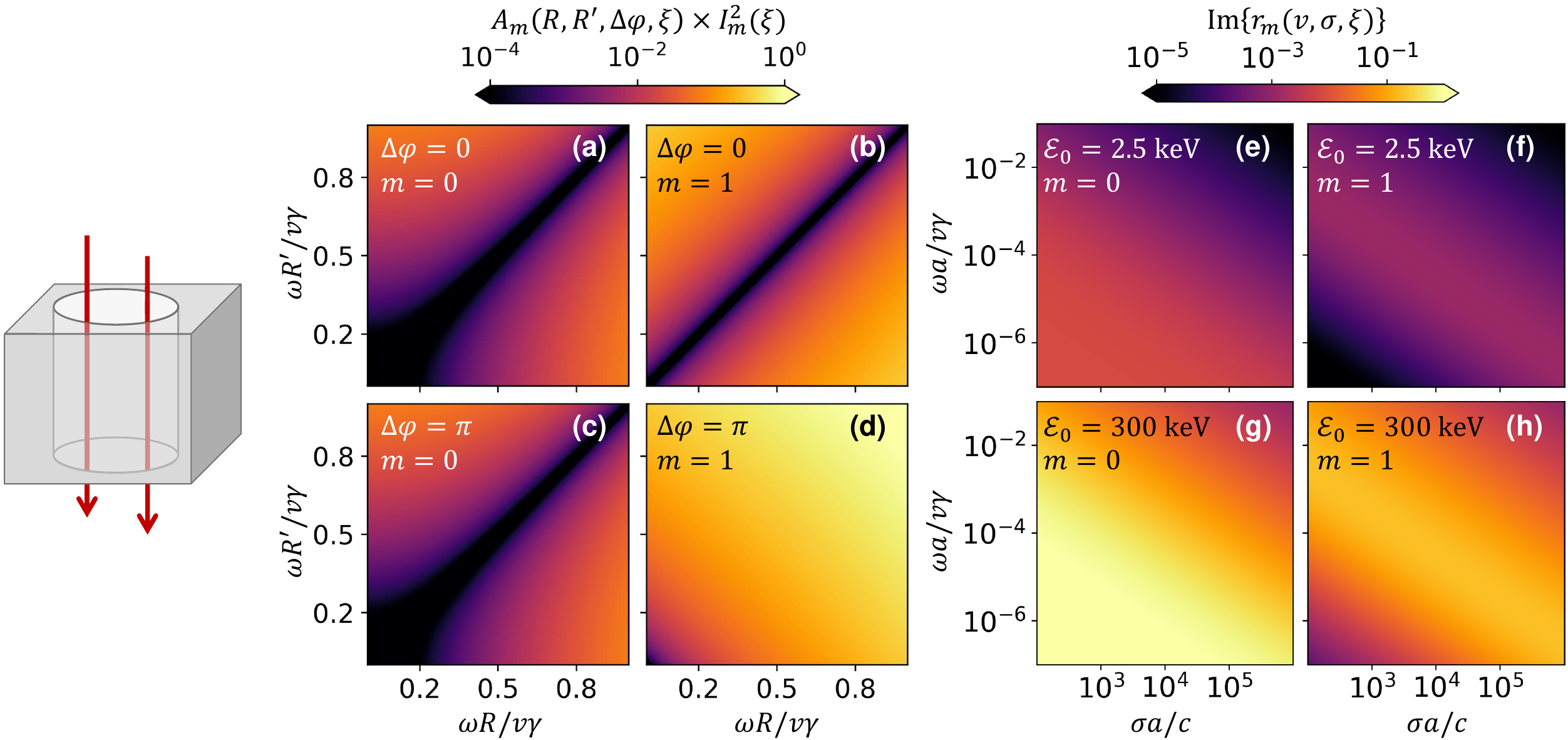}
\caption{\textbf{Radial dependence of the decoherence probability}. \textbf{(a-d)}~Position- and frequency-dependent factor $A_m(R,R',\Delta \varphi,\xi)$ [Eq.~(\ref{Am})], normalized by $I_m^2(\xi)$ and plotted as a function of $\xi R/a=\omega R/v\gamma$ and $\xi R'/a=\omega R'/v\gamma$ for $m=0,1$ and $\Delta \varphi=0,\pi$. \textbf{(e--h)}~Material-dependent contribution $\Imm\{r_m(v,\sigma,\xi)\}$ of the internal-hole reflection coefficient [Eq.~(\ref{rmsigma})], which enters the integrand of Eq.~(\ref{decoPres}), plotted as a function of $\xi=\omega a/v\gamma$ and $\sigma a/c$ for $m=0,1$ and electron kinetic energies $\mathcal{E}_0=2.5$ and $300$~keV.}
\label{Fig4}
\end{figure*}

The temperature, position, and conductivity dependences factorize within the contribution of each mode $m$ to the integrand of Eq.~(\ref{decoPres}). The path-position dependence is entirely encapsulated in the factor $A_m(R,R',\Delta\varphi,\xi)$, which we explore in Fig.~\ref{Fig4}(a--d) for different values of $R$, $R'$, and the angular separation $\Delta\varphi$. The positions are normalized by the frequency- and velocity-dependent factor $\omega/v\gamma$, which, after multiplying $A_m$ by $I_m^2(\xi)$, renders these plots universal.. Comparing the contributions of modes $m=0$ and $m=1$, we find that for $\Delta\varphi=0$ both behave qualitatively similarly, with the factor vanishing as $R=R'$. For $\Delta\varphi=\pi$, the $m=0$ mode retains this behavior, reflecting its independence of the azimuthal angle. At the same time, the $m=1$ contribution now vanishes only at $R=R'=0$, since the angular separation keeps the two paths distinguishable to the mode as long as they remain away from the center of the hole (and hence from each other). The material response is fully characterized by the factor derived from the reflection coefficient, $\Imm\{r_m(v,\sigma,\xi)\}$, shown in Fig.~\ref{Fig4}(e--h) as a function of the normalized frequency $\xi=\omega a/v\gamma$ and the normalized dc conductivity for different electron kinetic energies. For the $m=0$ mode, the factor grows as both the conductivity and the frequency decrease, whereas for $m=1$ it develops a localized maximum near $\omega\sim c^2/(a^2\sigma)$. In both cases, the factor increases with the electron kinetic energy, scaling as $\sim(v\gamma/c)^2$ and being markedly larger at $\mathcal{E}_0=300$~keV than at $2.5$~keV.

%----------------------------------------------------------
\subsection{High-conductivity limit} \label{highcond}

\begin{figure}[h]\centering\includegraphics[width=0.45\textwidth]{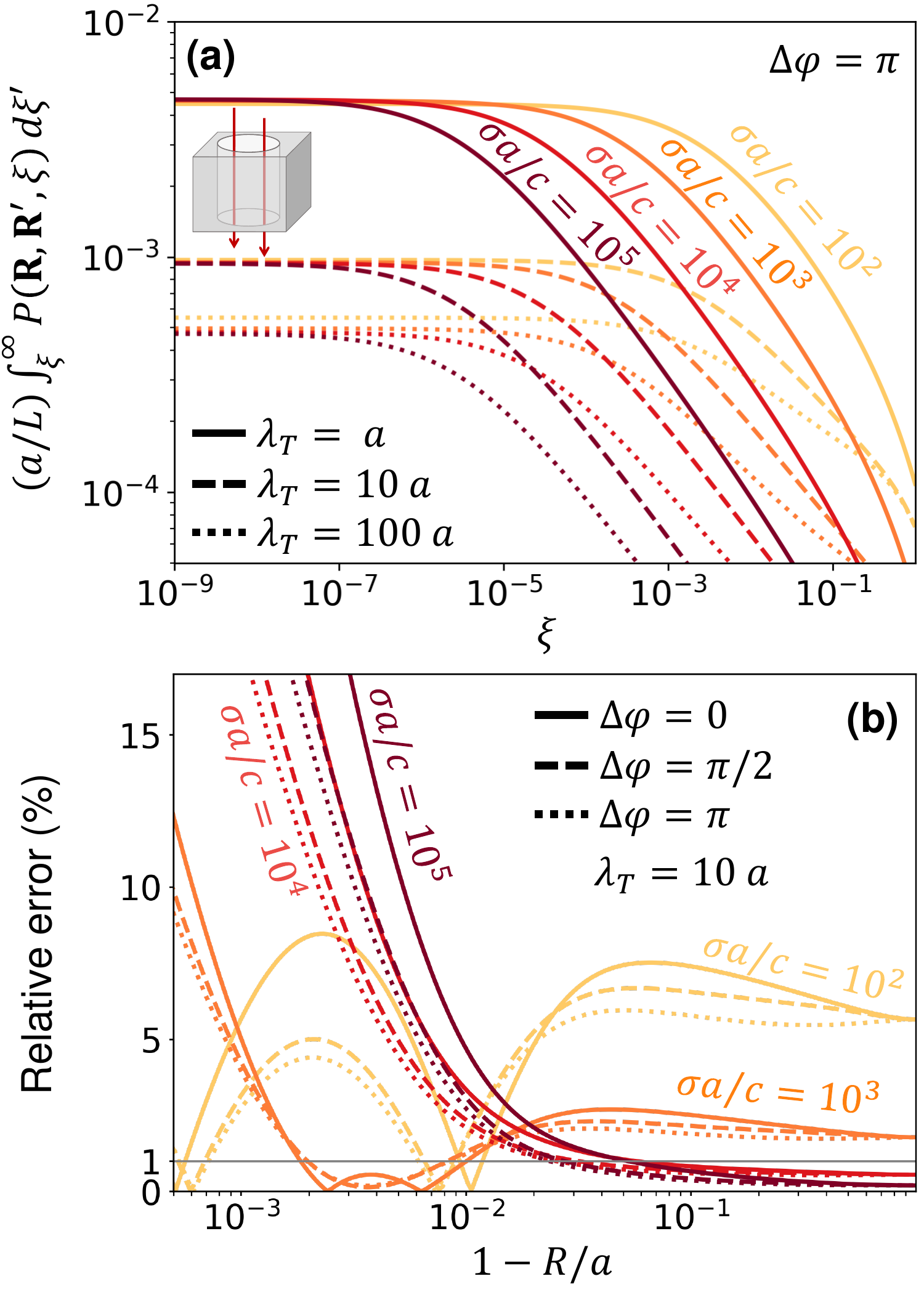}
\caption{\textbf{Low-frequency, high-conductivity limit.} \textbf{(a)}~Cumulative frequency integral of the decoherence-probability spectral density, $(a/L) \int_\xi^\infty P(\Rb,\Rb',\xi')\, d\xi'$, plotted as a function of the lower integration limit $\xi$ for different values of the dimensionless conductivity $\sigma a/c$ (colors) and thermal wavelength $\lambda_T$ (line styles), with $\Delta\varphi=\pi$ and $R=R'=0.5\,a$. \textbf{(b)}~Relative error of the approximation in Eq.~(\ref{approxP2}), plotted as a function of $1-R/a$ for different angular separations $\Delta \varphi$ (line styles) and values of $\sigma a/c$ [color-coordinated with panel (a)], with $\lambda_T=10\,a$ and $R'=0.5\,R$.}
\label{Fig5}
\end{figure}

As mentioned above, the conductivity factor $\sigma a/c$ is large for metallic materials and holes with radii in the range from $10\,{\rm \mu m}$ to $1\,{\rm mm}$. Indeed, using $c/\sigma=530$, $560$, $737$, and $884$~pm for silver, copper, gold, and aluminum, respectively~\cite{HLB17}, we find $\sigma a/c\sim10^5$ for $a\sim 50 \,{\rm \mu m}$. The frequency integral for the decoherence probability is then dominated by the low-frequency modes with $\xi\ll1$, as shown in Fig.~\ref{Fig5}(a). There we see that for $\sigma a/c=10^3$, the integrated probability is built up almost entirely from $\xi<10^{-2}$. It is then reasonable to approximate $f_m(\xi)\approx m/\xi+\mathcal{O}(\xi)$ and $\epsilon\approx4\pi\ii\sigma/\omega$, the latter giving $\xit\approx\sqrt{-4\pi\ii v\gamma \sigma a\xi}/c$, with which the conductivity-dependent factor becomes
%--------------------------
\begin{align}\nonumber%--
\Imm\{r_m(v,\sigma,\xi)\} \approx \bigg( \frac{v\gamma}{c} \bigg)^2 \Imm\bigg\{ \frac{1}{m-\xit g_m(\xit)} \bigg\}
\\\label{rmsmfr}%--
\approx\bigg( \frac{v\gamma}{c} \bigg)^2 \Imm\left\{\frac{K_m(\xit)}{\xit K_{m+1}(\xit)}\right\},
\end{align}
where the second line follows from the modified-Bessel recurrence $(m/x)K_m(x)-K_m'(x)=K_{m+1}(x)$. We do not low-frequency-expand $g_m(\xit)$, since $\xit\propto a/\delta_\omega$, where $\delta_\omega$ is the penetration depth introduced above. Over the relevant frequency range, we have $\delta_\omega\sim a$, so $\xit$ is not small.

% Figure 5 ------------------------------------------------

The remaining factors in Eq.~(\ref{decoPres}) simplify for $\xi\ll1$. In the temperature factor, given moderate values of $\lambda_T/a$, we can write
\begin{align}\nonumber%--
\coth\bigg(\frac{v\gamma\lambda_T}{4\pi a c}\,\xi\bigg) \approx \frac{4\pi a c}{v\gamma\lambda_T\xi},
\end{align}
which corresponds to a temperature boosting the population of modes with wavelengths of order $\lambda_T$ and larger. For the position factor $A_m(R,R',\Delta\varphi,\xi)$ [Eq.~(\ref{Am})], we have
\begin{align}\nonumber%--
&A_m(R,R',\Delta\varphi,\xi)
\\\nonumber%--
&\approx \left\{
\begin{array}{ll}
\xi^4 (R^2-R^{\prime2})^2/16a^4 & \quad\quad m = 0\\ & \\
(R/a)^{2m}+(R'/a)^{2m} & \\
\quad\quad -2\cos(m\Delta\varphi) (RR'/a^2)^{m} & \quad\quad m \geq 1
\end{array} \right.
\end{align}

The $m=0$ term is negligible because it is proportional to $\xi^4$, consistent with the results in Sec.~\ref{finitecond}, so we obtain
%--------------------------
\begin{widetext}%--
\begin{align}\label{approxP1}%--
P(\Rb,\Rb') = \frac{8\alpha L}{\lambda_T} \sum_{m=1}^{\infty} \left[\left(\frac{R}{a}\right)^{2m}+\left(\frac{R'}{a}\right)^{2m}-2\cos(m\Delta\varphi)\left(\frac{RR'}{a^2}\right)^{m}\right] \Imm\left\{ \int_0^{\infty} d\eta\, \frac{(1+\ii)K_m[(1-\ii)\eta]}{\eta^2 K_{m+1}[(1-\ii)\eta]}\right\},
\end{align}
\end{widetext}%--
where we have introduced the integration variable $\eta = \sqrt{2\pi v\gamma \sigma a\xi}/c = \xit/(1-\ii)$. This expression explicitly shows that, in the high-conductivity limit, the decoherence probability becomes independent of both conductivity and electron velocity. This weak material dependence was reported experimentally in Ref.~\cite{UMZ15} for an e-beam traveling through cylindrical tubes of different composition, where the decoherence was likewise found to be proportional to temperature, as in Eq.~(\ref{approxP1}).

The integral over $\eta$ can be evaluated in closed form by contour integration, giving
\begin{align}\label{contint}%--
\Imm\left\{ \int_0^{\infty} d\eta\, \frac{(1+\ii)K_m[(1-\ii)\eta]}{\eta^2 K_{m+1}[(1-\ii)\eta]}\right\} =\frac{\pi}{4m}
\end{align} 
(see Appendix~\ref{apdxproof} for the derivation). Introducing this into Eq.~(\ref{approxP1}), we find
\begin{align}\label{approxP2}%--
P(\Rb,\Rb')& = \frac{2\pi\alpha L}{\lambda_T}\ln\left[\frac{\big|a^2-\ee^{\ii\Delta\varphi}RR'\big|^2}{(a^2-R^2)(a^2-R^{\prime 2})}\right],
\end{align}
where we used the Taylor expansion $-\ln(1-x)=\sum_{m=1}^{\infty}x^m/m$. This approximation matches the exact result well, as shown in Fig.~\ref{Fig5}(b), where we plot the relative error $|P_{\rm exact}-P_{\rm approx}|/P_{\rm exact}$. The error stays below $1\%$ for $R/a<0.9$ when $\sigma a/c>10^4$. Away from the walls, the relative error decreases with conductivity, remaining below $5\%$ over a wide range of radii $R/a<0.99$ even for the relatively low value of $\sigma a/c=10^3$. Thus, for high conductivities, Eq.~(\ref{approxP2}) faithfully reproduces the exact decoherence probability unless the beam grazes the walls of the hole.

In this high-conductivity limit, the decoherence is given by the closed-form, universal expression of Eq.~(\ref{approxP2}), depending only on the path positions $(\Rb,\Rb')$, the geometric ratio $L/a$, and the temperature through $\lambda_T$, and independent of both the conductivity $\sigma$ and the electron velocity $v$. This velocity independence is nontrivial because retardation is fully retained in the underlying formulation and the independence emerges only in the high-conductivity, low-frequency limit. The logarithmic divergence of Eq.~(\ref{approxP2}) as either path approaches the wall ($R\to a$ or $R'\to a$) reflects the growing weight of high-order azimuthal modes there, while the linear scaling with temperature and with $L/a$ offers a simple estimate of the decoherence expected in a given experimental geometry.

%----------------------------------------------------------
\subsection{Comparison with results for a planar surface} \label{planar}

\begin{figure*}\centering\includegraphics[width=\textwidth]{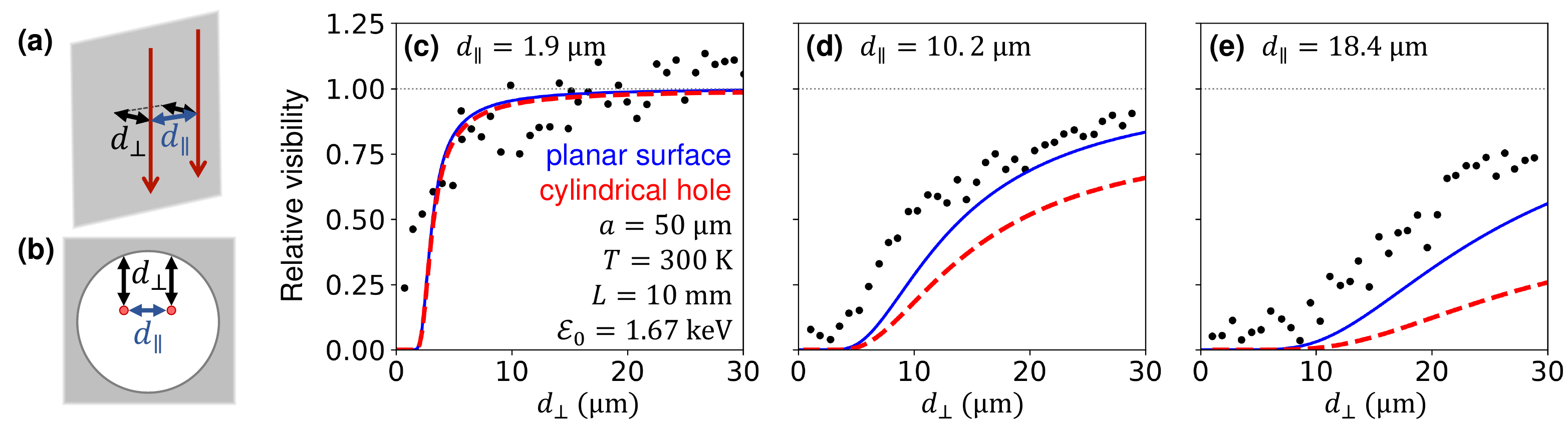}
\caption{\textbf{Comparison of decoherence in a cylindrical hole and a planar surface.} \textbf{(a)}~A two-path electron (interpath separation $d_\parallel$) moves with velocity $v$ parallel to the surface of an n-doped silicon slab. The two paths lie at a distance $d_\perp$ from the surface of the material, measured along the direction perpendicular to both the path separation and the electron velocity. \textbf{(b)}~Top-down view of a two-path electron moving inside a cylindrical hole of radius $a$ [see Fig.~\ref{Fig1}(a)] drilled through an n-doped silicon slab. As in panel (a), the paths are separated by a distance $d_\parallel$ and lie at a distance $d_\perp$ from the hole wall, measured along the direction perpendicular to both the electron velocity and the line segment joining the two paths. \textbf{(c)}~Relative visibility of the interference pattern as a function of $d_\perp$ for an electron traveling inside a cylindrical hole of radius $a=50\,\mu$m (red) and above a planar slab (blue), with $d_\parallel = 1.9\,\mu$m. The results are compared with the experimental data of Ref.~\cite{SH07} (black dots). The temperature, interaction length, and electron kinetic energy are fixed at $T=300$~K, $L=10$~mm, and $\mathcal{E}_0=1.67$~keV, respectively, to match the experimental conditions. The silicon permittivity is modeled using Eq.~(\ref{epsilonsi}).
\textbf{(d,e)}~Same as panel (c), but with $d_\parallel=10.2\,\mu$m and $18.4\,\mu$m, respectively. The visibility is computed as $\ee^{-P}$, where $P$ is the decoherence probability (see Sec.~\ref{twoslit} for details).}
\label{Fig6}
\end{figure*}

We now compare the cylindrical geometry studied here with the paradigmatic decoherence configuration in which a two-path electron travels parallel to the planar surface of a material slab, as sketched in Fig.~\ref{Fig6}(a). The two paths are separated by a distance $d_\parallel$ along the direction parallel to the surface and perpendicular to the electron velocity, while both remain at the same height $d_\perp$ above the surface. This geometry underlies the extensive theoretical and experimental work discussed in Sec.~\ref{Introduction}, including the measurements by Sonnentag and Hasselbach~\cite{SH07}. In that experiment, the relative visibility $\ee^{-P}$ of the interference pattern obtained by recombining the two paths (see Sec.~\ref{twoslit} for details) was measured as a function of $d_\perp$ for several fixed values of $d_\parallel$, following the interaction of the electron with an n-doped silicon slab. These data provide a natural benchmark for our results when both the interpath separation and the distance to the wall are small compared with the hole radius. In this regime, the curvature of the wall is negligible over the region sampled by the electron, and the decoherence induced by the cylindrical hole should approach that produced by a planar surface.

% Figure 6 ------------------------------------------------

To this end, we also compare our cylindrical calculation with the planar model of Ref.~\cite{SB12,paper464}, which treats the slab as an infinite half-space and yields a closed-form expression for the decoherence probability. When both paths travel in vacuum above the slab ($d_\perp>0$), the decoherence probability reads
\begin{widetext}%--
\begin{align} \label{decoplanar}%--
P^{{\rm planar}}(d_\parallel,d_\perp) &= \frac{2Le^2}{\pi\hbar v^2}\int_0^{\infty} d\omega [2n_T(\omega)+1]\int_0^\infty dq_\parallel\,\kappa_\perp\ee^{-2\kappa_\perp d_\perp}\,\frac{1-\cos(q_\parallel d_\parallel)}{q_\parallel^2+(\omega/v)^2} \,
 \Imm\left\{\left(\frac{v^2q_\parallel^2}{c^2\kappa_\perp^2} r_{\rm s} + r_{\rm p} \right)\right\}.
\end{align}    
\end{widetext}%--
Here, $q_\parallel$ denotes the in-plane wave-vector component along the interpath direction, while the component along the electron velocity is fixed to $\omega/v$ by the phase-matching condition. The quantity $\kappa_\perp = \sqrt{(\omega/v\gamma)^2+q_\parallel^2}$ is the decay constant of the evanescent field away from the surface. The Fresnel reflection coefficients of the vacuum--slab interface are $r_{\rm s}=(q_{\perp0}-q_{\perp1})/(q_{\perp0}+q_{\perp1})$ and $r_{\rm p}=(\epsilon q_{\perp0}-q_{\perp1})/(\epsilon q_{\perp0}+q_{\perp1})$, where $q_{\perp0}=\ii\kappa_\perp$ and $q_{\perp1}=\sqrt{k^2\epsilon-q_\parallel^2-(\omega/v)^2}$ are the normal wave-vector components in vacuum and inside the medium, respectively, with $\Ree\{q_{\perp1}\}>0$, $\Imm\{q_{\perp1}\}>0$, and $k=\omega/c$. The factor $1-\cos(q_\parallel d_\parallel)$ encodes the which-path information carried away by the material excitations: modes with wavelengths much larger than $d_\parallel$ cannot resolve the two trajectories and therefore do not contribute significantly to decoherence.

Because the sample is doped silicon rather than a good metal, we replace the dc-conductivity Drude response used in the preceding sections with a Drude model supplemented by a bound-charge background,
\begin{align} \label{epsilonsi}
\epsilon(\omega) = \epsilon_L \left[ 1- \frac{\omega_D^2}{\omega(\omega+\ii\gamma_D)} \right]
\end{align}
with $\epsilon_L = 11.7$, $\hbar\omega_{D} \approx 16$~meV and $\hbar\gamma_D \approx 1.7$~meV~\cite{KRS20,K26}. Here, $\omega_D$ is the screened plasma frequency and $\gamma_D$ is the damping rate. The parameter $\epsilon_L$ accounts for the lattice and interband polarizability of the silicon host, while the Drude term describes the free carriers introduced by doping. These carriers dominate the response at the sub-meV frequencies relevant here and are therefore not captured by tabulated optical constants of intrinsic silicon. These parameter values were extracted for the silicon surface studied in Ref.~\cite{KRS20}, whose interferometer shares many features with that of Ref.~\cite{SH07}, including the use of n-doped silicon samples. With these values, the comparison lies well outside the high-conductivity regime considered in Sec.~\ref{highcond}: for $a=50\,\mu$m, one obtains $\sigma a/c \approx 40$. Consequently, Eq.~(\ref{approxP2}) is inapplicable, and the full frequency-dependent calculation of Eq.~(\ref{decoPres}) is required. The response of the n-doped silicon is incorporated into the model presented in Sec.~\ref{Interaction} by setting $\sigma = (\ii\omega/4\pi)[1-\epsilon(\omega)]$, with $\epsilon(\omega)$ given by Eq.~(\ref{epsilonsi}).

We translate the planar coordinates ($d_\parallel$, $d_\perp$) into the cylindrical geometry by placing the two paths symmetrically about a radius. The paths are separated by a line segment of length $d_\parallel$, and their separation from the hole wall along the direction perpendicular to both the electron velocity and that segment is $d_\perp$ [Fig.~\ref{Fig6}(b)]. Their transverse coordinates are then $R=R'=\sqrt{y_0^2+(d_\parallel/2)^2}$ and $\Delta\varphi=2\arctan[d_\parallel/(2y_0)]$, where $y_0 = \sqrt{a^2-(d_\parallel/2)^2}-d_\perp$.

The resulting visibilities are shown in Fig.~\ref{Fig6}(c--e) for three interpath separations. For the smallest separation, $d_\parallel=1.9\,\mu$m [Fig.~\ref{Fig6}(c)], the two models are essentially indistinguishable over the full range of $d_\perp$, and both reproduce the measured visibility well. As $d_\parallel$ increases, the predictions begin to diverge [Fig.~\ref{Fig6}(d,e)], with the cylindrical geometry systematically producing stronger decoherence and, consequently, a smaller visibility.

Two distinct dependences govern this behavior. The ratio $P^{{\rm cyl}}/P^{{\rm planar}}$ quantifies the sensitivity of the electron to the wall curvature and is controlled primarily by $d_\perp/a$. The modes responsible for decoherence are long-wavelength modes, with the evanescent factor $\ee^{-2\kappa_\perp d_\perp}$ in Eq.~(\ref{decoplanar}) restricting the dominant contribution to $\kappa_\perp\lesssim1/2d_\perp$, corresponding roughly to $\lambda\sim10\,d_\perp$. When $d_\perp\ll a$, these modes probe only a portion of the wall that is small compared to the circumference of the hole, so the surface is locally indistinguishable from a plane. By contrast, when $d_\perp\sim a/2$, the relevant modes extend over an appreciable fraction of the hole. At $d_\perp=30\,\mu$m the ratio $P^{{\rm cyl}}/P^{{\rm planar}}$ is $2.27$, $2.27$, and $2.28$ in Fig.~\ref{Fig6}(c), (d), and (e), respectively. It therefore remains essentially unchanged even though $d_\parallel$ varies by nearly an order of magnitude. The interpath separation instead determines the overall magnitude of $P$, through the factor $1-\cos(q_\parallel d_\parallel)$ of Eq.~(\ref{decoplanar}) and its counterpart in the cylindrical mode sum [Eq.~(\ref{Am})], and does so almost equally in both geometries.

These two dependences act in opposition: $P$ decreases with increasing $d_\perp$, whereas the ratio $P^{{\rm cyl}}/P^{{\rm planar}}$ increases. Since the visibility is sensitive to this ratio mainly in the crossover region where $P\sim1$, the role of $d_\parallel$ is effectively to shift that crossover along the $d_\perp$ axis. For $d_\parallel=1.9\,\mu$m, the crossover occurs at $d_\perp\approx3\,\mu$m, where $d_\perp/a=0.06$ and the ratio is only $1.10$. The two visibility curves therefore remain close throughout the panel, with a maximum separation of $0.03$. For $d_\parallel=10.2$ and $18.4\,\mu$m the crossover shifts to $d_\perp\approx15$ and $28\,\mu$m, respectively, where the ratio has increased to $1.50$ and $2.17$. The corresponding maximum separations between the visibility curves are $0.17$ and $0.29$. The progression observed in Fig.~\ref{Fig6}(c--e) therefore reflects the displacement of the observable crossover region toward distances at which the curvature of the wall can no longer be neglected.

Finally, at the largest interpath separations, both models predict visibilities somewhat below the measured values. For the cylindrical model, this is consistent with the discussion above. For the planar model, two mechanisms could contribute. The first is instrumental: the surface position is difficult to determine precisely in these experiments~\cite{KRS20}. This effect alone, however, does not account for the full discrepancy, suggesting a second mechanism associated with finite-size effects: a sample of finite lateral extent suppresses modes with wavelengths comparable to or larger than the sample size, precisely the modes that become increasingly important as the beams are moved farther apart. 

In Supplementary Fig.~\ref{FigS2}, we repeat the calculation shown in Fig.~\ref{Fig6}(c--e) for an electron with kinetic energy $\mathcal{E}_0=80$~keV ($v\approx0.5\,c$), matching the velocity used elsewhere in this work. Although the electron is then approximately six times faster, the predicted visibilities remain nearly unchanged, differing from those in Fig.~\ref{Fig6}(c--e) by $\approx0.06$ at most over the entire plotted range in either geometry.

% =========================================================
\section{Decoherence by a circular aperture in a thin metallic film}
\label{film}

We now apply our formalism to the opposite geometric limit, $L\to0$, by considering an electron that crosses a circular aperture of radius $a$ in a thin metallic film, which we model as a perfect electric conductor (PEC, $\sigma\to\infty$), as sketched in Fig.~\ref{Fig1}(b). Besides describing a distinct and experimentally accessible configuration, this calculation captures a contribution absent from the infinite-hole treatment of Sec.~\ref{Interaction}: here the electron couples to radiative modes owing to the broken translational symmetry along the propagation direction. This mechanism, a close relative of transition radiation~\cite{GF1946} known as diffraction radiation~\cite{P98}, was shown to produce a long-range loss of coherence for a PEC half-plane~\cite{paper425} and can contribute significantly to decoherence~\cite{paper464}.

% Figure 7 ------------------------------------------------
\begin{figure*}\centering\includegraphics[width=1.0\textwidth]{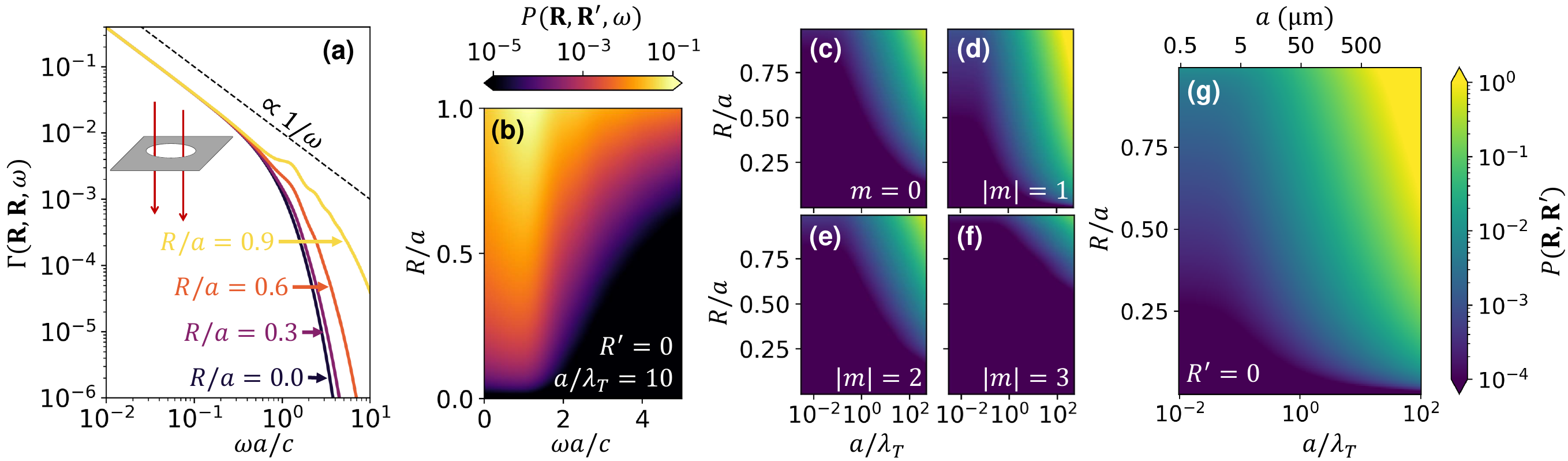}
\caption{\textbf{Decoherence induced by a circular aperture in a thin metallic film.} \textbf{(a)}~Single-path energy-loss probability $\Gamma_{\rm EELS}(R,\omega)\equiv\Gamma(\Rb,\Rb,\omega)$ for an electron traversing a circular aperture of radius $a$ in a PEC film [see Fig.~\ref{Fig1}(b,c)], plotted as a function of frequency for different radial positions $R/a$. The dashed line indicates the low-frequency $\propto1/\omega$ divergence. \textbf{(b)}~Spectral density of the decoherence probability for a two-path electron traversing the circular aperture [see Fig.~\ref{Fig1}(c)] with one path fixed at $R'=0$, plotted as a function of frequency and of the radial position $R$ of the second path for $a/\lambda_T=10$. \textbf{(c)}~Contribution of the azimuthal mode $m=0$ to the integrated decoherence probability for $v=0.5\,c$ and $R'=0$, plotted as a function of $R/a$ and $a/\lambda_T$. \textbf{(d--f)}~Same as panel (c), but for $|m|=1$, $|m|=2$, and $|m|=3$, respectively. \textbf{(g)}~Total decoherence probability obtained by summing over all azimuthal orders $m$. The upper horizontal axis gives the aperture radius for a temperature $T=300$~K ($\lambda_T\approx50\,\mu$m).}
\label{Fig7}
\end{figure*}

Because translational symmetry is broken, the aperture problem cannot be reduced to a single reflection coefficient. We instead solve Maxwell's equations subject to the PEC boundary conditions: the in-plane electric field vanishes on the film outside the aperture ($R>a$), whereas the in-plane electric and magnetic fields are continuous across the aperture ($R<a$). To this end, we expand the total field in propagating cylindrical waves on either side of the film and in standing cylindrical waves within the aperture plane. Projecting the boundary conditions onto this basis yields a linear system of equations for the modal amplitudes, which we solve numerically. The full derivation, including the explicit expressions for all matrix elements and for the coefficients describing the evanescent field generated by the moving electron, is presented in Appendix~\ref{apdxaperture}. In the subwavelength limit $\omega a/c\ll1$, our formalism recovers the classic Bethe--Bouwkamp description of diffraction by a small hole~\cite{B1944,B1941}, while retaining the full frequency dependence beyond that limit.

The solution provides the induced field through a set of cylindrical-wave amplitudes $\beta_{mQ}^{\nu}(\Rb',\omega)$, where $\nu={\rm s,p}$ labels the polarization, $m$ the azimuthal number, and $Q$ the transverse wave vector. The dependence on the electron path position $\Rb'=(R',\varphi')$ factors out as $\beta_{mQ}^{\nu}=(\ii e/v\gamma)\,\ee^{-\ii m\varphi'}I_m(\omega R'/v\gamma)\,\tilde\beta_{mQ}^{\nu}$, so that the aperture scattering is fully encoded in the path-independent reduced amplitudes $\tilde\beta_{mQ}^{\nu}$. It is then convenient to define the radial coefficient
\begin{align}\label{Bm}%--
B_m(R,\omega)=\int_0^{\infty}\frac{Q^2\,dQ}{Q^2+(\omega/v\gamma)^2}\,J_m(QR)\,\tilde\beta_{mQ}^{\rm p}(\omega),
\end{align}
in terms of which the generalized energy-loss probability, obtained by inserting the numerically computed induced field into the field-based expression of Eqs.~(\ref{Gammasym}) and (\ref{Gammatilde}), becomes
\begin{widetext}%--
\begin{align}\label{gammaaperture}%--
\Gamma(\Rb,\Rb',\omega)=\frac{e^2c}{\pi\hbar\omega v^2\gamma}\sum_m\Ree\bigg\{\ee^{\ii m\Delta\varphi}I_m\bigg(\frac{\omega R'}{v\gamma}\bigg)B_m(R,\omega)
+\ee^{-\ii m\Delta\varphi}I_m\bigg(\frac{\omega R}{v\gamma}\bigg)B_m(R',\omega)\bigg\}.
\end{align}
Only the induced (scattered) part of the field contributes to Eq.~(\ref{gammaaperture}), since the direct evanescent field of the electron cannot transfer energy in free space owing to the kinematic mismatch between the electron and free photons. Inserting Eq.~(\ref{gammaaperture}) into Eq.~(\ref{decoP}) yields the decoherence probability
\begin{align}\nonumber%--
P(\Rb,\Rb')=\frac{e^2c}{\pi\hbar v^2\gamma}\int_0^\infty\frac{d\omega}{\omega}\coth\bigg(\frac{\hbar\omega}{2k_BT}\bigg)\sum_m\Ree\bigg\{I_m\bigg(\frac{\omega R}{v\gamma}\bigg)B_m(R,\omega)+I_m\bigg(\frac{\omega R'}{v\gamma}\bigg)B_m(R',\omega)&
\\\label{Paperture}%--
-\ee^{\ii m\Delta\varphi}I_m\bigg(\frac{\omega R'}{v\gamma}\bigg)B_m(R,\omega)-\ee^{-\ii m\Delta\varphi}I_m\bigg(\frac{\omega R}{v\gamma}\bigg)B_m(R',\omega)&\bigg\},
\end{align}
\end{widetext}%--
where $\coth(\hbar\omega/2k_BT)=2n_T(\omega)+1$. An analogous expression for the elastic phase $\chi(\Rb,\Rb')$ is given in Appendix~\ref{apdxaperture}.

We begin by looking at the single-path energy-loss probability $\Gamma(\Rb,\Rb,\omega)$, which coincides with the usual electron energy-loss spectroscopy (EELS) probability~\cite{paper149} at $T=0$, shown in Fig.~\ref{Fig7}(a) as a function of $\omega a/c$ for several radial positions $R/a$. The loss probability diverges as $1/\omega$ in the low-frequency limit, mirroring the behavior found for a PEC half-plane~\cite{paper425} and reflecting the long-range nature of the diffraction-radiation coupling. As the path approaches the rim of the aperture ($R/a\to1$), the energy-loss probability develops oscillations, which arise from the growing weight of high-order azimuthal modes whose radial profiles peak near the edge.

The spectral density of the decoherence probability for a two-path electron is shown in Fig.~\ref{Fig7}(b) as a function of frequency and of the position $R$ of one path, with the other fixed at $R'=0$, for $a/\lambda_T=10$. Although $\Gamma(\Rb,\Rb,\omega)$ diverges as $1/\omega$ and the thermal factor contributes an additional $1/\omega$ at low frequency, the spectral density of $P$ peaks around $\omega a/c\sim1$ and saturates to a constant as $\omega\to0$. The infrared divergence is thus regularized in the decoherence probability because in the long-wavelength limit the two paths become indistinguishable to the field, so the bracketed combination $\Gamma(\Rb,\Rb,\omega)+\Gamma(\Rb',\Rb',\omega)-2\Gamma(\Rb,\Rb',\omega)$ in Eq.~(\ref{decoP}) vanishes and compensates the diverging prefactor, exactly as for the half-plane~\cite{paper425}.

The integrated decoherence probability is presented in Fig.~\ref{Fig7}(c--g), following a similar scheme to Fig.~\ref{Fig2}(b--f) for the cylindrical hole. Panels (c--f) show the contributions of the $m=0$ and $|m|=1$--$3$ modes, respectively, and panel (g) shows the result of summing over all $m$, as functions of the radial position and of $a/\lambda_T$. As in the cylindrical case, the decoherence probability grows with temperature, owing to the thermal population of low-frequency modes, and with path separation, since well-separated paths leave more distinguishable traces in the field. The qualitative dependence closely parallels that of the $L/a$-normalized probability of the cylindrical hole, with the notable difference that the $m=0$ contribution is now appreciable, comparable to that of $|m|=2$. In addition, the increase in the decoherence probability as $R$ approaches $a$ is smoother in the circular aperture considered in Fig.~\ref{Fig7}. The $m=0$ mode is negligible for the cylinder, where the high conductivity confines the decoherence to low frequencies and the $m=0$ term is suppressed as $\xi^4$, the same suppression that justifies neglecting it in the high-conductivity limit (see Sec.~\ref{highcond}). For the aperture, by contrast, the radiative coupling shifts the dominant spectral weight to $\omega a/c\sim1$, where the $m=0$ mode is no longer suppressed and is comparable to the higher-order contributions. The dominant contribution nevertheless remains that of the $|m|=1$ modes.

There is, however, a quantitative difference. Because the aperture interaction is localized, $P$ does not grow with any interaction length, and its magnitude is therefore bounded, in contrast to the cylindrical-hole result, which scales linearly with $L$. Moreover, while configurations such as an electron passing a semi-infinite plane allow the path separation to be made arbitrarily large~\cite{paper149,paper371}, here it is capped by the aperture radius, which limits the attainable decoherence. As a consequence, the aperture contribution is negligible compared with that of a long cylindrical hole whenever $L\gg a$, and becomes appreciable only for large apertures at high temperature (i.e., large $a/\lambda_T$), where thermally boosted high-order modes dominate. These observations justify neglecting the aperture contribution in the analysis of image formation presented in Sec.~\ref{imaging}, where $L\gg a$.

% =========================================================
\section{The effect of decoherence on image formation}
\label{imaging}

Having characterized the decoherence induced by the two geometries, we now examine how it manifests in observable quantities relevant to electron microscopy. We focus on a cylindrical hole drilled through a slab of finite conductivity and thickness $L\gg a$, using the decoherence probability $P(\Rb,\Rb')$ and elastic phase $\chi(\Rb,\Rb')$ derived in Sec.~\ref{Interaction}, and neglecting the localized contributions from the entrance and exit apertures, which are small in this regime (see Sec.~\ref{film}).

To assess the effects of decoherence on the e-beam, we consider two experimental methods for probing the beam's coherence: a two-slit experiment and the image formed by focusing the beam with an electron lens. In both cases, the relevant optical element (i.e., the slits or the lens) is placed immediately after the beam exits the hole at $z=z_0$. Here, its state is
\begin{align}\nonumber%--
\rho(\Rb,z_0,\Rb',z_0)= t(\Rb) t^{*}(\Rb') \ee^{-P(\Rb,\Rb')+\ii\chi(\Rb,\Rb')}\rho_i(\Rb,\Rb'),
\end{align}
where $t(\Rb)$ is the transmission factor of the slits or lens and 
\begin{align}\nonumber%--
\rho_i(\Rb,\Rb') = \frac{1}{\pi b^2}\,\Theta(b-R)\,\Theta(b-R')
\end{align}
is the initial state of the electron before traversing the hole. The latter is taken to be a uniform beam of radius $b$. The state at a later position $z_1$ is~\cite{paper451}
\begin{align}\nonumber%--
\rho(\Rb,z_1,\Rb',z_1)
=\int \frac{d^2\Qb}{(2\pi)^2}&\int \frac{d^2\Qb'}{(2\pi)^2}\; \ee^{\ii (\Qb\cdot\Rb-\Qb'\cdot\Rb')}
\\\nonumber%--
& \times \ee^{\ii (q_z-q_z')d}\; \rho(\Qb,z_0,\Qb',z_0),
\end{align}
where $d=z_1-z_0$, $q_z=\sqrt{q_0^2-Q^2}$, and $\hbar q_0=mv\gamma$ is the e-beam's momentum. The quantity $\rho(\Qb,z_0,\Qb',z_0)=\int d^2\Rb \int d^2\Rb'\;\ee^{-\ii (\Qb\cdot\Rb-\Qb'\cdot\Rb')} \rho(\Rb,z_0,\Rb',z_0)$ is the momentum-space representation of the beam's density matrix immediately after the hole. At a screen located at $z_1$, the beam intensity is $I(\Rb,d) = j_0\,  \rho(\Rb,z_1,\Rb,z_1)$, with $j_0$ the incident current density.

%----------------------------------------------------------
\subsection{Two-slit experiment} \label{twoslit}

We consider a screen of negligible thickness that blocks the electron except at two circular apertures of radius $w$ at points $\Rb_1$ and $\Rb_2$ in the transverse plane, for which the transmission factor is
\begin{align}\nonumber%--
t(\Rb) = \frac{1}{\sqrt{2}} \Big[\Theta(w-|\Rb-\Rb_1|)+\Theta(w-|\Rb-\Rb_2|)\Big],
\end{align}
where $\Theta(x)$ is equal to $1$ for $x>0$ and $0$ otherwise. When the hole radius $a$ and the slit separation $|\Rb_1-\Rb_2|$ are much larger than the slit radius $w$, the decoherence probability is approximately constant across each slit, so that
$\ee^{-P(\Rb+\Rb_1,\Rb'+\Rb_2)} \approx \ee^{-P(\Rb_1,\Rb_2)}$. The resulting intensity at the screen is then
\begin{align}\label{doubleslit}%--
I(\Rb,d) =& j_0\Big[h_{11}(\Rb,d)+h_{22}(\Rb,d)
\\\nonumber%--
&+2\ee^{-P(\Rb_1,\Rb_2)} \Ree\{\ee^{\ii\chi(\Rb_1,\Rb_2)}h_{12}(\Rb,d)\}\Big],
\end{align}
where $h_{ij}(\Rb,z) = (1/2\pi)\,g(|\Rb-\Rb_i|,z) g^{*}(|\Rb-\Rb_j|,z)$ with
\begin{align}\nonumber%--
g(x,z) = \frac{w}{b} \int_0^{\infty} dQ \ee^{\ii q_z z} J_1(Qw) J_0(Q x).
\end{align}
The first and second terms on the right-hand side of Eq.~(\ref{doubleslit}) represent the sum of the intensities from each slit, while the third term represents interference. As expected, when different spatial parts of the beam lose coherence, the interference term vanishes exponentially, as the decoherence probability at the slit centers $P(\Rb_1,\Rb_2)$ increases.

% Figure 8 ------------------------------------------------
\begin{figure*}\centering\includegraphics[width=1.0\textwidth]{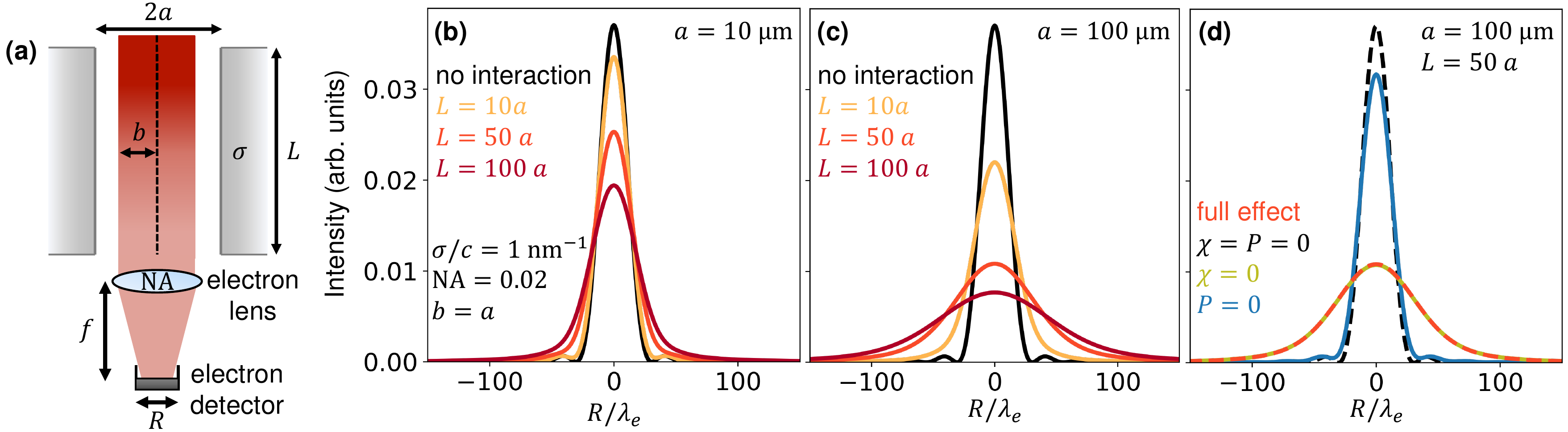}
\caption{\textbf{Effects of interaction on image formation.} \textbf{(a)}~A broad e-beam of width $b$ propagates through a hole of radius $a$ drilled in a slab of conductivity $\sigma$ and thickness $L$. A converging electron lens of radius $b$ and focal length $f$ (numerical aperture ${\rm NA}=b/f=0.02$) is placed immediately at the exit of the hole, and an electron analyzer in the focal plane is displaced by a distance $R$ from the optical axis. \textbf{(b,c)}~Intensity at the analyzer for a beam that fills the hole ($b=a$), plotted as a function of the analyzer displacement $R$ normalized by the electron wavelength $\lambda_e$, for different slab thicknesses $L$ (colors), with \textbf{(b)}~$a=10\,\mu$m and \textbf{(c)}~$a=100\,\mu$m. The black line corresponds to the case without interaction. \textbf{(d)}~Analyzer intensity for $a=100\,\mu$m and $L=50\,a$, comparing the full calculation with results obtained by selectively neglecting the interaction-induced phase $\chi$ or decoherence probability $P$, as well as with the noninteracting case. In all panels, $v=0.5\,c$, $T=300$~K, ${\rm NA}=0.02$, and $\sigma/c=1\,{\rm nm}^{-1}$. The intensity profiles for different values of $L$ are normalized such that $2\pi\int_0^{\infty}R\,dR\,I(R,f)=1$.}
\label{Fig8}
\end{figure*}

%----------------------------------------------------------
\subsection{Focusing by a lens}

Now we consider the system represented in Fig.~\ref{Fig8}(a), in which a convergent electron lens with a focal length $f$ and radius $b$ equal to that of the beam, characterized by its numerical aperture ${\rm NA} \approx b/f$, is placed at the end of the hole. For this system, the transmission factor is given by
\begin{align}\nonumber%--
t(R) = \ee^{-\ii q_0R^2/2f}\Theta(b-R).
\end{align}
For a lens with a small numerical aperture, we take the paraxial approximation, $q_z \approx q_0 - Q^2/2q_0$, in which case the intensity at the focal plane ($d=f$) is
\begin{align}\nonumber%--
I(\Rb,f) = \int_{R'<b} d^2\Rb' \int_{R''<b} d^2\Rb''\; \ee^{-\ii (q_0/f) \Rb \cdot(\Rb'-\Rb'')}&
\\\nonumber%--
\times\frac{j_0\, q_0^2}{4\pi^3 (fb)^2}\;\ee^{-P(R',R'',\varphi'-\varphi'')+\ii\chi(R',R'')}&. 
\end{align}
Owing to rotational invariance, the intensity is independent of the observation angle, so one of the two azimuth integrals can be carried out in closed form. To do so, we introduce the angular separation $\varphi = \varphi'-\varphi''$ as an integration variable and rewrite $\Rb \cdot(\Rb'-\Rb'')$ as $R A \cos(\varphi''+\delta)$, with the radial amplitude $A=\sqrt{R'^2+R''^2-2R'R''\cos\varphi}$ and an offset $\delta$ that can be readily absorbed by $\varphi''$. This allows us to use the Bessel integral representation $2\pi J_0(x)=\int_0^{2\pi}d\varphi''\,\ee^{\ii x\cos\varphi''}$. The intensity thus reads
\begin{widetext}%--
\begin{align}\label{intlens}%--
I(R,f) &= \frac{2 j_0 f^2}{(\lambda_e b)^2} \int_0^{2\pi} d\varphi \int_0^{\rm NA} \theta d\theta \int_0^{\rm NA} \theta' d\theta'
 \ee^{-P(\theta f,\theta'f,\varphi)+\ii \chi(\theta f,\theta' f)}
 J_0\bigg(2\pi \frac{R}{\lambda_e}\sqrt{\theta^2+\theta^{\prime 2}-2\theta\theta^{\prime} \cos\varphi}\bigg),
\end{align}
\end{widetext}%--
where $\theta=R'/f$, $\theta'=R''/f$, and $\lambda_e=2\pi/q_0$ is the electron wavelength.

The changes in intensity for different hole parameters are shown in Fig.~\ref{Fig8}(b--c), where Eq.~(\ref{intlens}) is plotted as a function of the analyzer position $R$ for different values of $L/a$ when the beam fills the hole ($b=a$). As the effects of the interaction become more pronounced, the characteristic Airy pattern grows more diffuse as the central peak broadens and loses intensity, with the effect being stronger for larger hole radii. In Fig.~\ref{Fig8}(d) we compare how the phase and the decoherence probability independently affect the focal spot. Decoherence dominates, particularly for large holes, where retaining the decoherence probability alone (and neglecting the phase) provides a good approximation to the full result. Conversely, neglecting decoherence and retaining only the phase produces only modest changes in the focused intensity profile, leaving a well-focused image. For completeness, we present analogous results for the circular aperture of Sec.~\ref{film} in Supplementary Fig.~\ref{FigS3}. The corresponding effects are negligible when compared to those shown in Fig.~\ref{Fig8}.

% =========================================================
\section{Conclusions} 
\label{conclusion}

We have developed a rigorous theory of free-electron decoherence arising from electromagnetic interactions with two canonical conducting geometries. Using a Green-tensor formulation combined with the fluctuation--dissipation theorem, we first considered an electron propagating through a long cylindrical hole in a conductive material, with the hole length much greater than its radius. The resulting azimuthally resolved expression for the decoherence probability provides a transparent decomposition into contributions from different frequency ranges and azimuthal orders, while separating the dependence on trajectory position, temperature, and conductivity. In the high-conductivity and high-temperature regime, the response is dominated by low-frequency modes and the decoherence probability reduces to a universal expression that is independent of both the conductivity and the electron velocity, and increases linearly with temperature.

To establish contact with experiment, we also studied the standard configuration of a two-path electron moving above a planar surface. We compared the predictions of the planar and cylindrical geometries with the interferometric measurements of Sonnentag and Hasselbach~\cite{SH07} and found good agreement at the smallest interpath separations $d_\parallel$. The comparison isolates the effect of curvature in the cylindrical hole, which becomes significant when the wavelengths of the dominant modes are comparable to or larger than the hole radius.

We then examined the complementary geometry of an electron traversing a circular aperture in a thin metallic film, solving the corresponding boundary-value problem using a cylindrical-wave modal expansion. The energy-loss probability diverges as $1/\omega$ at low frequency, as expected from the perfectly conducting half-plane geometry~\cite{paper425}, whereas the decoherence probability remains finite and has its dominant spectral weight near $\omega a/c\sim1$. Because the aperture interaction is spatially localized, its contribution is much smaller than the decoherence accumulated along a long cylindrical hole when $L\gg a$. It becomes appreciable only for sufficiently large apertures with a radius comparable to or larger than the thermal wavelength ($a/\lambda_T\gtrsim1$), where the weight of high-order modes is increased by the thermal population.

Finally, we quantified the influence of decoherence on interference and image formation in two-slit and electron-lens configurations. For sufficiently large holes, decoherence becomes the dominant mechanism degrading a focused electron probe, producing substantially greater broadening than the interaction-induced elastic phase alone. In particular, the enhancement found for the enclosed geometry implies that estimates based on a planar surface systematically underestimate the coherence lost by a beam traveling inside a hole. These results establish environment-induced decoherence from nearby conducting structures as an effect that must be considered in the design and interpretation of coherent electron-beam experiments and instruments.

The theory presented here can be readily adapted to arbitrary transverse electron wave-function profiles and specimen geometries. However, when the interaction distance is large, lateral spreading of the transverse wave function becomes important. A complete description would then involve tracking the evolution of the electron density matrix along the beam-propagation direction in a piecewise manner. Specifically, the propagation could be divided into a sequence of intervals, each long enough for the corresponding decoherence contribution to be well defined, but short enough that lateral spreading remains negligible within that interval. The evolution would then be obtained by successively applying the factor $\ee^{-P+\ii\chi}$ appearing in Eq.~(\ref{evol}) over each interval, together with the lateral spreading discussed elsewhere~\cite{paper371}.

% =========================================================
\acknowledgments

We are deeply indebted to Archie Howie for many stimulating and enjoyable discussions and for generously sharing his insights, which encouraged us to explore the effect of decoherence on the focused electron intensity and to compare our theory with experimental data. This work was partially supported by the European Research Council (101141220-QUEFES) and the Spanish MICINN (PID2020-112625GB-I00 and Severo Ochoa CEX2019-000910-S).

\appendix
% =========================================================
\section{Loss probability for an electron inside a cylindrical hole}
\label{apdxeels}

For an electron traveling along the cylindrical hole described above, the induced Green tensor follows from the free-space Green tensor [the solution of Eq.~(\ref{greentens}) with $\epsilon = 1$],
\begin{align}\label{greenfree}%--
&G_{zz}^0(\rb-\rb',\omega) = \frac{-1}{4\pi\omega^2} (k^2+\partial_{zz}^2) \frac{\ee^{\ii k |\rb-\rb'|}}{|\rb-\rb'|} 
\\\nonumber%--
&= \int \frac{dq}{2\pi} \ee^{\ii q(z-z')} \sum_m \frac{\kappa^2}{2\pi\omega^2} \ee^{\ii m (\varphi-\varphi')} I_m(\kappa R_<) K_m(\kappa R_>),
\end{align}
where $\rb=(R,\varphi,z)$ denotes cylindrical coordinates, $R_> = \max\{R,R'\}$, $R_< = \min\{R,R'\}$, $\kappa = \sqrt{q^2-k^2-\ii 0^{+}}$ (with $\Ree\{\kappa\}>0$ and $\Imm\{\kappa\}<0$), and $k=\omega/c$. The second line of Eq.~(\ref{greenfree}) follows from the Sommerfeld identity, $\exp(\ii kr)/r = (1/\pi)\int dq \ee^{\ii q z} K_0(\kappa R)$, together with Graf's theorem, $ K_0(\kappa |\Rb-\Rb'|) = \sum_m K_m(\kappa R_{>})I_m(\kappa R_{<}) \ee^{\ii m (\varphi-\varphi')}$ [see Eq.~(9.1.79) in Ref.~\cite{AS1972}]. Only the Green tensor evaluated inside the hole, along the electron path, is required. In this region, the modified Bessel function $K_m(\kappa R_>)$ in Eq.~(\ref{greenfree}) represents the $E_z$ component of an outgoing ${\rm p}$-polarized cylindrical wave generated by the electron's evanescent field, as discussed below. Reflection at the hole wall $R=a$ converts the outgoing wave into a standing wave with $I_m(\kappa R)$ radial dependence, giving
\begin{align}\label{gind}%--
G_{zz}^{\rm ind}(\rb,\rb',\omega) &= \int\frac{dq}{2\pi} \ee^{\ii q(z-z')} \frac{\ii \kappa^2}{4\omega^2}
\\\nonumber%--
&\times \sum_m \ee^{\ii m(\varphi-\varphi')} (-1)^m r^{m}_{{\rm pp}} I_m(\kappa R) I_m(\kappa R'),
\end{align}
which holds for $R,R'<a$. Here, $r_{{\rm pp}}^m$ is the internal ${\rm p}$-to-${\rm p}$ reflection coefficient. Substituting Eq.~(\ref{gind}) into Eq.~(\ref{eels}) and performing the $z$ integrals gives
\begin{align}\nonumber%--
\Gamma(\Rb,\Rb',\omega) =& \frac{e^2 L}{\hbar v^2 \gamma^2} \sum_{m=-\infty}^{\infty} I_m\bigg(\frac{\omega R}{v\gamma} \bigg) I_m\bigg(\frac{\omega R'}{v\gamma} \bigg)
\\\nonumber%--
&\times \Ree\big\{(-1)^{m+1} r_{{\rm pp}}^m \ee^{\ii m(\varphi-\varphi')}\big\},
\end{align}
where the factor $L$ (the hole length) arises from translational symmetry along $z$. This symmetry enforces energy--momentum conservation, yielding the phase-matching condition $q =\omega/v$, and hence, $\kappa =\omega/v\gamma$.

% =========================================================
\section{Derivation of the internal hole reflection coefficient} \label{refcoeff}

To describe electromagnetic waves in the cylindrical hole, we expand the field in a basis of cylindrical waves following Ref.~\cite{paper047}. We work in cylindrical coordinates $\rb = (R,\varphi,z)$ and decompose the fields in cylindrical waves $\Eb_{j,qm\nu}^{Z\pm}(\rb) = \ee^{\pm \ii qz+\ii m \varphi} \tilde{\Eb}_{j,qm\nu}^{Z\pm}(\Rb)$, with radial profiles
\begin{subequations}\label{cylbasis}%--
\begin{align}%--
\tilde{\Eb}_{j,qm{\rm s}}^{Z\pm}(\Rb) =& \frac{\ii m}{Q_j R} Z_m(Q_j R) \hat{\Rb} - Z_m'(Q_j R) \hat{\bm \varphi}
\\%--
\tilde{\Eb}_{j,qm{\rm p}}^{Z\pm}(\Rb) =& \pm\frac{q}{k_j} \bigg[ \ii  Z_m'(Q_j R) \hat{\Rb}
\\\nonumber%--
&- \frac{m}{Q_j R} Z_m(Q_j R) \hat{\bm \varphi}\pm \frac{Q_j}{q} Z_m(Q_j R) \hat{{\bf z}} \bigg],
\end{align}    
\end{subequations}
where $q$ is the wave vector component along $z$, $m$ is the azimuthal number, $\nu={\rm s},{\rm p}$ labels the polarization, and $j$ denotes the medium ($j = 0$ for the vacuum inside the hole and $j = 1$ for the surrounding material of permittivity $\epsilon$). Waves with $Z=J$ ($Z=H$) are regular (outgoing) waves, involving the Bessel function $J_m$ (Hankel function $H_m^{(1)}$), while $\pm$ labels the propagation direction of the waves along $z$. Finally, $k_j = \sqrt{\epsilon_j}\,\omega/c$ and $Q_j = \sqrt{k_j^2-q^2+\ii0^{+}}$, with the square-root branches chosen such that $\Ree\{Q_j\}>0$ and $\Imm\{Q_j\}>0$. 

We consider an outgoing wave $\Eb_{0,qm\nu}^H(\rb)$ emanating from inside the hole. Upon reflection at the wall $R=a$, this wave produces standing waves $\Eb_{0,qm\nu}^J$, so that the field inside the hole becomes
\begin{align}\nonumber%--
\Eb(\rb) = \Eb_{0,qm\nu + }^{H+}(\rb)+r_{s\nu}^{m}\Eb_{0,qm{\rm s}}^{J+}(\rb)+r_{p\nu}^{m}\Eb_{0,qm{\rm p}}^{J+}(\rb),
\end{align}
where $r_{\nu'\nu}^{m}$ are the sought-after reflection coefficients. In the surrounding material $j=1$, the field consists of transmitted outgoing waves,
\begin{align}\nonumber%--
\Eb(\rb) = t_{s\nu}^{m}\Eb_{1,qm{\rm s}}^{H+}(\rb)+t_{p\nu}^{m}\Eb_{1,qm{\rm p}}^{H+}(\rb),
\end{align}
where $t_{\nu'\nu}^m$ are the corresponding transmission coefficients. Enforcing continuity of the tangential electric and magnetic field components at the interface, after some lengthy but straightforward algebra, we find
\begin{align}\nonumber%--
\begin{bmatrix}
r_{\rm ss}^m\\
t_{\rm ss}^m \\
r_{\rm ps}^m\\
t_{\rm ps}^m
\end{bmatrix} = M^{-1} 
\begin{bmatrix}
(-Q_0/k_0) H_m^{(1)}(Q_0a)\\
-H_m^{(1)\prime}(Q_0a) \\
0\\
(-mq/k_1Q_0a) H_m^{(1)}(Q_0a)
\end{bmatrix}
\end{align}
and
\begin{align}\nonumber%--
\begin{bmatrix}
r_{\rm sp}^m\\
t_{\rm sp}^m \\
r_{\rm pp}^m\\
t_{\rm pp}^m
\end{bmatrix} = M^{-1} 
\begin{bmatrix}
0\\
(-mq/k_0Q_0a)H_m^{(1)}(Q_0a) \\
(-Q_0/k_0)H_m^{(1)}(Q_0a)\\
(-k_0/k_1) H_m^{(1)\prime}(Q_0 a)
\end{bmatrix},
\end{align}
where
\begin{widetext}%--
\begin{align}\nonumber%--
M = \begin{bmatrix}
(Q_0/k_0)J_m(Q_0 a) && (-Q_1/k_0) H_m^{(1)}(Q_1 a) && 0 && 0 \\
J_m'(Q_0 a) && -H_m^{(1)\prime}(Q_1 a) && (mq/k_0Q_0a) J_m(Q_0 a) && (-mq/k_1Q_1a) H_m^{(1)}(Q_1 a) \\
0 && 0 && (Q_0/k_0) J_m(Q_0 a) && (-Q_1/k_1) H_m^{(1)}(Q_1 a) \\
(mq/k_1 Q_0 a) J_m(Q_0 a) && (-mq/k_1 Q_1 a) H_m^{(1)}(Q_1 a) && (k_0/k_1) J_m'(Q_0 a) && -H_m^{(1)\prime}(Q_1 a)
\end{bmatrix}.
\end{align}
\end{widetext}%--
For an electron moving with velocity $\vb$ parallel to the hole axis, $q = \omega/v$, $Q_0 = \ii\omega/v\gamma$, and $Q_1 = \ii (\omega/v)\sqrt{1-(v/c)^2\epsilon}$, so the Bessel and Hankel functions convert to modified Bessel functions via the identities $J_m(\ii x) = \ii^m I_m(x)$ and $H_m^{(1)}(\ii x) = (2\ii^{-m-1}/\pi ) K_m(x)$, yielding Eq.~(\ref{rppm}).

% =========================================================
\section{Proof of Eq.~(\ref{contint})} \label{apdxproof}

We give a self-contained derivation of Eq.~(\ref{contint}): the imaginary part of $\int_0^{\infty} F(\eta) \,d\eta$, with
\begin{align}\nonumber%--
F(\eta)  = \frac{(1+\ii) K_m[(1-\ii)\eta]}{\eta^2 K_{m+1}[(1-\ii)\eta]}.
\end{align} 
To this end, we close the real-axis integration path using the contour shown in the following sketch:
\begin{center}\includegraphics[width = 0.25\textwidth]{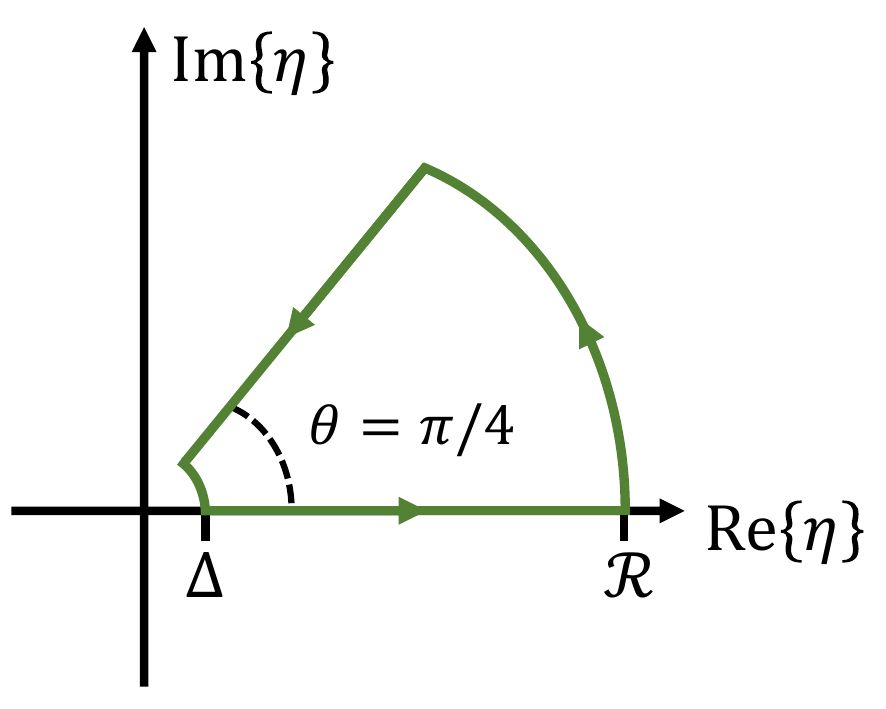}\end{center}%--
The contour is defined over the sector $0\leq\arg\{z\}\leq\pi/4$ and bounded by $\Delta \leq|z|\leq \mathcal{R}$, with the limits $\Delta\to0$ and $\mathcal{R}\to\infty$ taken at the end. The origin must be excluded because $K_m(z)/K_{m+1}(z)\sim z/2m$ as $z\to0$ for $m\neq0$, so that $F(\eta)$ has a simple pole there. Moreover, $K_{m+1}(z)$ has no zeros in the sector $|\arg\{z\}|\leq\pi/2$~\cite{AS1972}. Since the argument $(1-\ii)z$ lies between $-\pi/4$ and 0 throughout the contour and therefore remains in the zero-free sector, $F$ has no poles inside the contour. Cauchy's theorem then gives
\begin{align}\nonumber%--
&\lim_{\Delta\to0}\,\int_{\Delta}^{\infty} F(\eta) d\eta \\
&\quad= \lim_{\Delta\to0}\left[-\int_{\mathcal{A}_\Delta} F(z) dz + \ee^{\ii\pi/4}\int_{\Delta}^{\infty} F(\ee^{\ii\pi/4}x) dx \right],
\end{align}
where the limit $\mathcal{R}\to\infty$ has been taken, so that the outer-arc contribution vanishes since $F(z)\sim z^{-2}$ as $|z|\to\infty$. The inner-arc integral, with $z=\Delta\ee^{\ii\theta}$ traversed clockwise from $\theta=\pi/4$ to $\theta=0$, is
\begin{align}\label{contarc}%--
\lim_{\Delta\to0}\int_{\mathcal{A}_\Delta} F(z) dz = -\frac{\ii}{m}\int_{0}^{\pi/4} d\theta = -\frac{\ii \pi}{4m}.
\end{align}
In addition, the integral along the ray $\arg\{z\}=\pi/4$ develops a logarithmic divergence as $\Delta \to0$. However, this divergence is purely real. Indeed, along this ray one may write $z=x\ee^{\ii\pi/4}$, so that the argument of the modified Bessel functions becomes real ($(1-\ii)z = \sqrt{2}x$), while the remaining phase factors combine into a real prefactor. Consequently, the entire integrand along the ray is real. Taking the imaginary part therefore eliminates this contribution, leaving only the arc term in Eq.~(\ref{contarc}), from which Eq.~(\ref{contint}) follows.

% =========================================================
\section{Numerical solution for the circular aperture}
\label{apdxaperture}

We give here the modal scheme used to compute the field induced when the evanescent field of a moving electron interacts with a circular aperture of radius $a$ in a PEC film occupying the plane $z=0$, closely following Ref.~\cite{paper187}. Throughout this appendix, the standing-wave modal amplitudes $\alpha_{m\ell}^{\nu\pm}$ introduced below (always mode-indexed) are not to be confused with the fine-structure constant $\alpha$ of the main text.

%----------------------------------------------------------
\subsection{Field expansion}

We decompose the field into cylindrical waves $\Eb_{qm\nu}^{Z\pm}(\rb)=\ee^{\pm\ii qz+\ii m\varphi}\tilde{\Eb}_{qm\nu}^{Z\pm}(\Rb)$, with radial profiles given by Eq.~(\ref{cylbasis}). 
Since all fields here are in vacuum, we omit the medium index ($j=0$ throughout). The total electric field is written as $\Eb(\rb,\omega)=\Eb^{\rm ext}(\rb,\omega)+\Eb^{\rm ind}(\rb,\omega)$, where $\Eb^{\rm ext}$ is the external electron field, and
\begin{align}\nonumber%--
\Eb^{\rm ind}(\rb,\omega) =
\\\nonumber%--
\sum_{\nu m}&\left\{
\begin{matrix}
\displaystyle\int_0^{\infty}QdQ\,\beta_{mQ}^{\nu-}\,\Eb_{q_Qm\nu}^{J-}(\rb), & z<0\\[2.2ex]
\displaystyle\sum_{\ell,\pm}\alpha_{m\ell}^{\nu\pm}\,\Eb_{q_{Q_{m\ell\nu}}m\nu}^{J\pm}(\rb), & z=0\\[2.2ex]
\displaystyle\int_0^{\infty}QdQ\,\beta_{mQ}^{\nu+}\,\Eb_{q_Qm\nu}^{J+}(\rb), & z>0
\end{matrix}\right.
\end{align}
is the induced part. The regions $z\neq0$ contain outward-propagating waves with continuous transverse wave vector $Q$, while the field in the aperture plane ($z=0$), which serves as a matching plane, is expanded in standing waves with discrete radial wave vectors $Q_{m\ell\nu}$ fixed by $J_m(aQ_{m\ell{\rm p}})=0$ and $J_m'(aQ_{m\ell{\rm s}})=0$. Here, the $z$ component of the wave vector is given in terms of $Q$ as $q_Q = \sqrt{k^2-Q^2+\ii 0^{+}}$ (with $\Ree\{q_Q\}>0$ and $\Imm\{q_Q\}>0$). The magnetic field follows from Faraday's law,
\begin{align}\label{totfldH}%--
\Hb^{\rm ind}(\rb,\omega)&=
\\\nonumber%--
-\ii\sum_{\nu m}&\left\{
\begin{matrix}
\displaystyle\int_0^{\infty}QdQ\,\beta_{mQ}^{\nu-}\,\Eb_{q_Qm\tilde\nu}^{J-}(\rb), & z<0\\[2.2ex]
\displaystyle\sum_{\ell,\pm}\alpha_{m\ell}^{\nu\pm}\,\Eb_{q_{Q_{m\ell\nu}}m\tilde\nu}^{J\pm}(\rb), & z=0\\[2.2ex]
\displaystyle\int_0^{\infty}QdQ\,\beta_{mQ}^{\nu+}\,\Eb_{q_Qm\tilde\nu}^{J+}(\rb), & z>0
\end{matrix}\right.
\end{align}
where $\tilde\nu$ denotes the conjugate polarization ($\tilde{\rm s}={\rm p}$ and vice versa).

%----------------------------------------------------------
\subsection{Boundary conditions and linear system of equations}

The PEC aperture imposes the following conditions:
\renewcommand{\theenumi}{\alph{enumi}}
\begin{enumerate} 
\item continuity of the parallel electric field across $z=0$ for all $\Rb$;
\item continuity of the parallel magnetic field across the aperture, $R<a$;
\item vanishing parallel electric field on the film, $R>a$.
\end{enumerate}
Condition (a) relates the amplitudes on the two sides of the film, $\beta_{mQ}^{{\rm p}-}=-\beta_{mQ}^{{\rm p}+}$ and $\beta_{mQ}^{{\rm s}-}=\beta_{mQ}^{{\rm s}+}$, and likewise, $\alpha_{m\ell}^{{\rm p}-}=-\alpha_{m\ell}^{{\rm p}+}$ and $\alpha_{m\ell}^{{\rm s}-}=\alpha_{m\ell}^{{\rm s}+}$. We now project the fields onto the basis of cylindrical waves and apply the orthogonality relation
\begin{align}%--
\int_0^{\infty}R\,dR\,\big[\tilde{\Eb}_{q_{Q'}m\nu'}^{J+}(\Rb)\big]_\parallel^*\cdot\tilde{\Eb}_{q_Qm\nu}^{J\pm}(\Rb)
\\\nonumber%--
=\delta_{\nu\nu'}\frac{\delta(Q-Q')}{Q}N_{Q\nu}^\pm,
\end{align}
where $N_{Q{\rm p}}^\pm=\pm|q_Q|^2/k^2$ and $N_{Q{\rm s}}^\pm=1$. We also need the overlap of standing and propagating waves over the aperture,
\begin{align}%--
\int_0^{a}R\,dR\,\left[\tilde{\Eb}_{q_Qm\nu'}^{J+}(\Rb)\right]_\parallel^*\cdot\tilde{\Eb}_{q_{Q_{\ell\nu}}m\nu}^{J\pm}(\Rb)=S_{\ell Q\nu'\nu}^{\pm},
\end{align}
whose nonzero elements are
\begin{align}\nonumber%--
&S_{\ell Q{\rm pp}}^{\pm}=\pm\frac{q_{Q_{\ell{\rm p}}}q_Q^{*}}{k^2}\,I(Q_{\ell{\rm p}},Q),
\\\nonumber%--
&S_{\ell Q{\rm ss}}^{\pm}=I(Q,Q_{\ell{\rm s}}),
\\\nonumber%--
&S_{\ell Q{\rm ps}}^{\pm}=\frac{q_Q^{*}m}{kQQ_{\ell{\rm s}}}J_m(Qa)J_m(Q_{\ell{\rm s}}a)
\end{align}
in terms of the auxiliary function
\begin{align}\nonumber%--
I(Q,Q')=\frac{Q'a}{Q'^2-Q^2}J_m'(Qa)J_m(Q'a).
\end{align}
Using these results, conditions (a) and (c) combine into
\begin{align}\label{conda}%--
&\sum_{\nu'\ell\pm}\alpha_{m\ell}^{\nu'\pm}S_{\ell Q\nu\nu'}^{\pm} 
\\\nonumber%--
&=N_{Q\nu}^+\beta_{mQ}^{\nu+}+\frac{1}{2\pi}\int_{R>a}d^2\Rb\,\big[\tilde{\Eb}_{q_Qm\nu}^{J+}(\Rb)\big]^*_\parallel\cdot\Eb^{\rm ext}(\Rb),
\end{align}
where condition (c) has been used to limit the external-field projection to the region $R>a$. Projecting the magnetic field [Eq.~(\ref{totfldH})] onto the conjugate standing-wave basis over the aperture introduces the integrals
\begin{align}\nonumber%--
\int_0^{a}R\,dR\,\left[\tilde{\Eb}_{q_{Q_{\ell'\nu'}}m\tilde\nu'}^{J+}\right]_\parallel^*\cdot\tilde{\Eb}_{q_{Q_{\ell\nu}}m\tilde\nu}^{J\pm}=\delta_{\nu\nu'}\delta_{\ell\ell'}M_{\ell\nu}^{\pm},
\end{align}
with
\begin{align}\nonumber%--
M_{\ell{\rm p}}^{\pm}&=(a^2/2)J_{m+1}^2(Q_{\ell{\rm p}}a)
\\\nonumber%--
M_{\ell{\rm s}}^{\pm}&=\pm\frac{1}{2}\left|\frac{aq_{Q_{\ell{\rm s}}}}{k}\right|^2J_m^2(Q_{\ell{\rm s}}a)\left[1-\left(\frac{m}{Q_{\ell{\rm s}}a}\right)^2\right]
\end{align}
and the mixed overlaps $\tilde{S}_{\ell Q\nu\nu'}^{\pm}$ with nonzero elements
\begin{subequations}%--
\begin{align*}%--
&\tilde{S}_{\ell Q{\rm pp}}^{\pm}=I(Q_{\ell{\rm p}},Q), \\
&\tilde{S}_{\ell Q{\rm ss}}^{\pm}=\pm\frac{q_{Q_{\ell{\rm s}}}^*q_Q}{k^2}I(Q,Q_{\ell{\rm s}}), \\
&\tilde{S}_{\ell Q{\rm sp}}^{\pm}=\frac{mq_{Q_{\ell{\rm s}}}^*}{kQQ_{\ell{\rm s}}}J_m(Qa)J_m(Q_{\ell{\rm s}}a).
\end{align*}
\end{subequations}
Condition (b) then reduces to
\begin{align}\label{condb}%--
\int_0^{\infty}Q\,dQ\,\sum_{\nu'}\tilde{S}_{\ell Q\nu\nu'}^{+}\beta_{mQ}^{\nu'+}=0.
\end{align}
Isolating $\beta_{mQ}^{\nu+}$ from Eq.~(\ref{conda}) and inserting it into Eq.~(\ref{condb}) yields a linear system of equations for the standing-wave amplitudes:
\begin{subequations}\label{eqalphabeta}%--
\begin{align}%--
&\sum_{\nu'\ell'}A_{\ell\ell'}^{\nu\nu'}\alpha_{m\ell'}^{\nu'+}=b_{\ell\nu},
\\%--
&\beta_{mQ}^{\nu+}=\frac{1}{N_{Q\nu}^+}\bigg[2\sum_{\nu'\ell}S_{\ell Q\nu\nu'}^{+}\alpha_{m\ell}^{\nu'+}-C_{mQ}^\nu\bigg],
\end{align}
\end{subequations}
where the system matrix is
\begin{align}\nonumber%--
A_{\ell \ell'}^{\nu \nu'} = 2 \int_0^{\infty} Q dQ \sum_{\nu''} \frac{\tilde{S}^+_{\ell Q \nu \nu''} S^+_{\ell' Q \nu'' \nu'}}{N_{Q\nu''}^+}.
\end{align}
The dependence on the external field enters only through
\begin{subequations}%--
\begin{align}\label{ccoef}%--
&C_{mQ}^\nu=\frac{1}{2\pi}\int_{R>a}d^2\Rb\,\left[\Eb_{q_Qm\nu}^{J+}(\Rb)\right]_\parallel^*\cdot\Eb^{\rm ext}(\Rb),
\\\label{bcoef}%--
&b_{\ell\nu}=\int_0^{\infty}QdQ\sum_{\nu'}\tilde{S}_{\ell Q\nu\nu'}^{+}\frac{C_{mQ}^{\nu'}}{N_{Q\nu'}^+}.
\end{align}
\end{subequations}
In practice, the integrals over $Q$ are discretized on a quadrature grid and the azimuthal index $m$ and standing-wave index $\ell$ are truncated until convergence is achieved.

%----------------------------------------------------------
\subsection{External field of the moving electron}

For a point electron moving along $\zz$ at transverse position $\Rb_i=(R_i,\varphi_i)$, the evanescent field for $R>R_i$ reads
\begin{align}\label{extfld}%--
\Eb^{\rm ext}(\rb,\omega)=&\frac{\ii e\pi\omega}{v\gamma c}\sum_{m=-\infty}^{\infty}\ee^{\ii m(\pi/2-\varphi_i)}
\\\nonumber%--
&\times I_m\bigg(\frac{\omega R_i}{v\gamma}\bigg)\Eb_{\omega/v,m,{\rm p}}^{H+}(\rb),
\end{align}
which consists of a superposition of outgoing ${\rm p}$-polarized cylindrical waves. Inserting Eq.~(\ref{extfld}) into Eq.~(\ref{ccoef}) reduces $C_{mQ}^\nu$ to projections of the outgoing field onto the regular basis waves over the film. The projection for $\nu={\rm s}$ polarization is straightforward, while the $\nu={\rm p}$ projection follows from Eqs.~6.521-2 and 6.521-4 of Ref.~\cite{GR1980}, which combine into
\begin{align}\nonumber%--
\int_a^{\infty}\!\! x\,&dx\,K_m(\mu x)J_m(\lambda x)=\frac{a}{\lambda^2+\mu^2}
\\\nonumber%--
&\times\big[\mu J_m(\lambda a)K_{m+1}(\mu a)-\lambda J_{m+1}(\lambda a)K_m(\mu a)\big].
\end{align}
The electron position factors out as $C_{mQ}^{\nu}=(\ii e/v\gamma)\,\ee^{-\ii m\varphi_i}I_m(\omega R_i/v\gamma)\,\tilde{C}_{mQ}^{\nu}$, with
\begin{widetext}%--
\begin{subequations}%--
\begin{align*}%--
\tilde{C}_{mQ}^{\rm p}&=\frac{q_Q^{*}c}{v[Q^2+(\omega/v\gamma)^2]}\bigg\{\frac{a\omega}{v\gamma}\Big[J_{m-1}(Qa)K_m\Big(\tfrac{\omega a}{v\gamma}\Big)-J_{m+1}(Qa)K_{m+2}\Big(\tfrac{\omega a}{v\gamma}\Big)\Big]
\\\nonumber
&\qquad\qquad\qquad\qquad-Qa\Big[J_{m}(Qa)K_{m-1}\Big(\tfrac{\omega a}{v\gamma}\Big)-J_{m+2}(Qa)K_{m+1}\Big(\tfrac{\omega a}{v\gamma}\Big)\Big]\bigg\},\\
\tilde{C}_{mQ}^{\rm s}&=\frac{2m\gamma}{Q}J_m(Qa)K_m\Big(\frac{\omega a}{v\gamma}\Big).
\end{align*}
\end{subequations}
\end{widetext}%--
Because $b_{\ell\nu}$ is linear in $C_{mQ}^{\nu'}$ [see Eq.~(\ref{bcoef})], we can factorize the position dependence out in the same way, and define the reduced amplitudes $\tilde\alpha_{m\ell}^{\nu}$ and $\tilde\beta_{mQ}^{\nu}$. Restoring the prefactor gives
\begin{align}\nonumber%--
\beta_{mQ}^{\nu+}(\Rb_i,\omega)=\frac{\ii e}{v\gamma}\,\ee^{-\ii m\varphi_i}I_m\bigg(\frac{\omega R_i}{v\gamma}\bigg)\,\tilde\beta_{mQ}^{\nu+},
\end{align}
where
\begin{align}\nonumber%--
\tilde\beta_{mQ}^{\nu+}=\frac{1}{N_{Q\nu}^+}\bigg[2\sum_{\nu'\ell}S_{\ell Q\nu\nu'}^{+}\tilde\alpha_{m\ell}^{\nu'}-\tilde{C}_{mQ}^{\nu}\bigg].
\end{align}
In practice, this means that for a given velocity the linear system [Eqs.~(\ref{eqalphabeta})] needs only be solved once at each frequency, after which the coefficients apply to any transverse position of the e-beam.

%----------------------------------------------------------
\subsection{Observable quantities}

The induced field within either half-space is $\Eb^{\rm ind}(\rb,\Rb_i,\omega)=\int_0^{\infty}QdQ\sum_{\nu m}\beta_{mQ}^{\nu\pm}(\Rb_i,\omega)\Eb_{q_Qm\nu}^{J\pm }(\rb)$. Projecting its $z$ component onto the electron trajectory and symmetrizing in $(\Rb,\Rb')$ yields the nonlocal energy-loss probability of Eq.~(\ref{gammaaperture}) in terms of the radial coefficient $B_m(R,\omega)$ of Eq.~(\ref{Bm}). The decoherence probability $P(\Rb,\Rb')$ [Eq.~(\ref{Paperture})] then follows through Eq.~(\ref{decoP}). The elastic phase is obtained analogously as
\begin{align}\nonumber%--
\chi(\Rb,\Rb')=\int_0^{\infty}d\omega\,\big[\phi(\Rb',\omega)-\phi(\Rb,\omega)\big],
\end{align}
where
\begin{align}\nonumber%--
\phi(\Rb,\omega)=\frac{e^2c}{\pi\hbar\omega v^2\gamma}\sum_m I_m\bigg(\frac{\omega R}{v\gamma}\bigg)\Imm\big\{B_m(R,\omega)\big\}
\end{align}
is a single-path phase.

% =========================================================
%\bibliography{refsL.bib,refs2.bib}
%\bibliography{../../../bibtex/refsL.bib}


% =========================================================
\pagebreak \onecolumngrid \section*{SUPPLEMENTARY FIGURES}
\renewcommand{\thefigure}{S\arabic{figure}}
\setcounter{figure}{0}

% Figure S1 ------------------------------------------------
\begin{figure*}[h!]\centering\includegraphics[width =\textwidth]{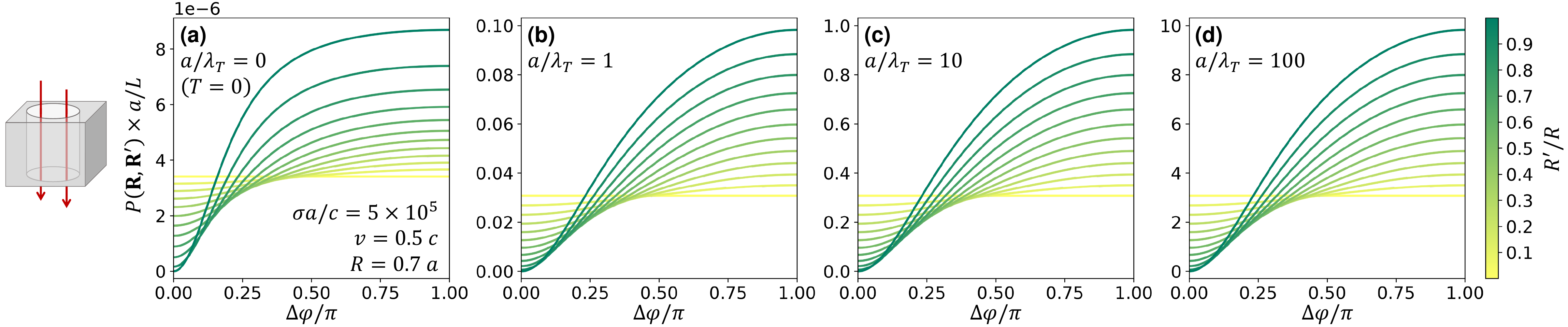}
\caption{\textbf{Effect of temperature on the angular dependence of the decoherence probability}. \textbf{(a)}~Decoherence probability, normalized by the hole length-to-radius ratio $L/a$ (i.e., $P\times a/L$), for electron paths located at polar coordinates $(R,\varphi)$ and $(R',\varphi')$ [see Fig.~1(c) of the main text], plotted as a function of their relative azimuthal separation $\Delta\varphi=\varphi-\varphi'$. Results are shown for $v=0.5\,c$, $\sigma a/c=5\times10^5$, $a/\lambda_T=0$ ($T=0$), $R=0.7\,a$, and several values of the $R'/R$ ratio (see color scale). \textbf{(b,c,d)}~Same as panel (a), but for $a/\lambda_T=1$, $10$, and $100$, respectively. Panel (c) corresponds to Fig.~3 of the main text.}
\label{FigS1}
\end{figure*}

% Figure S2 ------------------------------------------------
\begin{figure*}[h!]\centering\includegraphics[width =\textwidth]{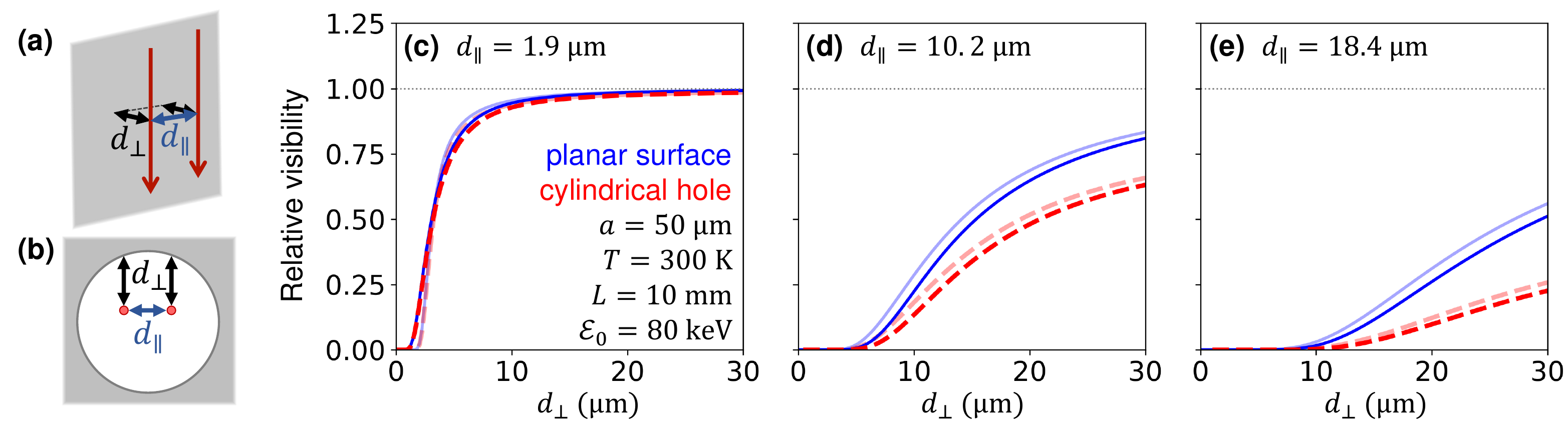}
\caption{\textbf{Comparison of decoherence in a cylindrical hole and above a planar surface for high-energy electrons.} We plot the visibility calculated under the same conditions as in Fig.~6 of the main text, but for electrons with a kinetic energy of 80~keV (color curves). For comparison, the results shown in Fig.~6 for 1.67~keV electrons are replotted in a lighter hue of the corresponding color. \textbf{(a)}~A two-path electron, with interpath separation $d_\parallel$, moves with velocity $v$ parallel to the surface of an n-doped silicon slab. The two paths lie at a distance $d_\perp$ from the surface, measured along the direction perpendicular to both the path separation and the electron velocity. \textbf{(b)}~Top-down view of a two-path electron moving inside a cylindrical hole of radius $a$ [see Fig.~1(a)] drilled through an n-doped silicon slab. As in panel (a), the paths are separated by a distance $d_\parallel$ and lie at a distance $d_\perp$ from the hole wall, measured along the direction perpendicular to both the electron velocity and the line segment joining the two paths. \textbf{(c)}~Relative visibility of the interference pattern as a function of $d_\perp$ for an electron traveling inside a cylindrical hole of radius $a=50\,\mu$m (red) and above a planar slab (blue), with $d_\parallel=1.9\,\mu$m. The temperature, interaction length, and electron kinetic energy are fixed at $T=300$~K, $L=10$~mm, and $\mathcal{E}_0=80$~keV, respectively. \textbf{(d,e)}~Same as panel (c), but for $d_\parallel=10.2\,\mu$m and $18.4\,\mu$m, respectively. The visibility is computed as $\ee^{-P}$, where $P$ is the decoherence probability. This figure is the high-kinetic-energy counterpart of Fig.~6 in the main text.}
\label{Fig4}
\end{figure*}

% Figure S3 ------------------------------------------------
\begin{figure*}[h!]\centering\includegraphics[width = 0.75\textwidth]{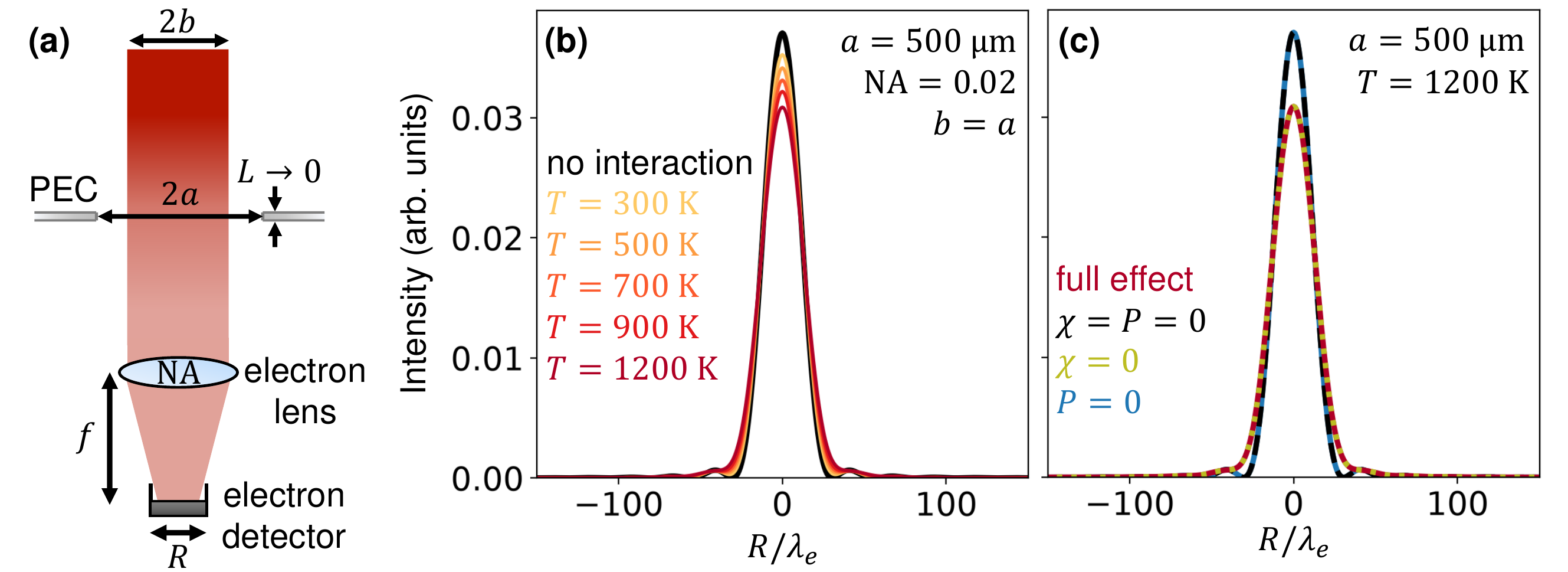}
\caption{\textbf{Effects of interaction with a circular aperture on image formation.} \textbf{(a)}~A broad e-beam of width $b$ propagates through an aperture of radius $a=500\,\mu$m in a zero-thickness PEC screen. A converging electron lens of radius $b$, focal length $f$, and numerical aperture ${\rm NA}=b/f=0.02$ is placed immediately after the screen, and an electron analyzer in the focal plane is displaced by a distance $R$ from the optical axis. \textbf{(b)}~Intensity at the analyzer for a beam that fills the hole ($b=a$), plotted as a function of the analyzer displacement $R$, normalized by the electron wavelength $\lambda_e$, for different temperatures $T$ (colors). The black curve corresponds to the noninteracting case. \textbf{(d)}~Analyzer intensity for $a=500\,\mu$m and $T=1200$~K, comparing the full calculation with results obtained by selectively neglecting either the interaction-induced phase $\chi$ or the decoherence probability $P$, as well as with the noninteracting case. In all panels, $v=0.5\,c$ and ${\rm NA}=0.02$. The intensity profiles for different values of $T$ are normalized such that $2\pi\int_{-\infty}^{\infty}R\,dR\,I(R,f)=1$.}
\label{FigS3}
\end{figure*}

\end{document}